\documentclass[11pt,letterpaper]{article}
\pdfoutput=1
\usepackage{jcappub}

\usepackage{appendix}
\usepackage{array}
\newcolumntype{x}[1]{>{\centering\arraybackslash}p{#1}}

\usepackage{epsfig}
\usepackage{graphicx}
\usepackage{dcolumn}
\usepackage{amsmath}
\usepackage{enumerate}
\usepackage{mathrsfs}

\usepackage{color}
\usepackage{ulem}
\hypersetup{colorlinks=false}

\newcommand{\eg}{e.g.~}
\newcommand{\ie}{i.e.~}

\newcommand{\beq}{\begin{equation}}
\newcommand{\eeq}{\end{equation}}
\newcommand{\ud}{\text{d}}

\newcommand{\mDM}{m}
\newcommand{\ER}{E_\text{R}}
\newcommand{\Ed}{E'}
\newcommand{\vesc}{v_\text{esc}}
\newcommand{\vmin}{v_\text{min}}

\newcommand{\bfv}{\mathbf{v}}
\newcommand{\eH}{\mathcal{H}}
\newcommand{\eR}{\mathcal{R}}

\newcommand{\spazio}{\bigskip}

\newcommand{\Lag}{\mathscr{L}}	

\usepackage{color}
\definecolor{rossoCP3}{cmyk}{0,.88,.77,.40}
\definecolor{verdeCP3}{rgb}{0.09765625, 0.57421875, 0.1015625}
\definecolor{bluCP3}{rgb}{0, 0.23, 0.67}

\ifx\pdfoutput\undefined
\usepackage[dvips,bookmarks]{hyperref}    
\else
\usepackage{hyperref}    
\fi
\hypersetup{colorlinks, bookmarksopen, bookmarksnumbered,
citecolor=verdeCP3, linkcolor=bluCP3, pdfstartview=FitH, urlcolor=rossoCP3}

\title{Halo-independent comparison of direct dark matter detection data: a review}

\author{Eugenio Del Nobile}

\affiliation{Department of Physics and Astronomy, UCLA,\\
475 Portola Plaza, Los Angeles, CA 90095, USA}

\emailAdd{delnobile@physics.ucla.edu}

\abstract{We review the halo-independent formalism, that allows to compare data from different direct dark matter detection experiments without making assumptions on the properties of the dark matter halo. We apply this method to spin-independent WIMP-nuclei interactions, for both isospin-conserving and isospin-violating couplings, and to WIMPs interacting through an anomalous magnetic moment.}

\begin{document}

\maketitle

\section{Introduction}
The presence of dark matter (DM) in the universe is now an established fact, that has been confirmed once more by the recent precise measurements of the Planck satellite \cite{Ade:2013zuv}. Many different particle candidates exist as possible explanations for the DM. A particular class of candidates, the WIMPs (for weakly interacting massive particles), is very actively searched for. WIMPs are particles with weakly interacting cross sections and masses in the 1 GeV/$c^2$ -- 10 TeV/$c^2$ range. Of particular interest are light WIMPs, with mass around 1 -- 10 GeV/$c^2$.

At present, four direct dark matter search experiments (DAMA \cite{Bernabei:2010mq}, CoGeNT \cite{Aalseth:2010vx, Aalseth:2011wp, Aalseth:2012if}, CRESST-II \cite{Angloher:2011uu}, and CDMS-II-Si \cite{Agnese:2013rvf}) have data that may be interpreted as signals from DM particles in the light WIMPs range. DAMA \cite{Bernabei:2010mq} and CoGeNT \cite{Aalseth:2011wp} report annual modulations in their event rates, compatible with those expected for a DM signal \cite{Drukier:1986tm, Freese:2012xd}. CoGeNT \cite{Aalseth:2010vx, Aalseth:2012if}, CRESST-II \cite{Angloher:2011uu}, and CDMS-II-Si \cite{Agnese:2013rvf}, observe an excess of events above their expected backgrounds, that may be interpreted as due to DM WIMPs.

However, other experiments do not observe significant excesses above their estimated background, thus setting upper limits on the interaction of WIMPs with nuclei. The most stringent limits on the average (unmodulated) rate for light WIMPs are set by the LUX \cite{Akerib:2013tjd}, XENON10 \cite{Angle:2011th}, XENON100 \cite{Aprile:2012nq}, CDMS-II-Ge \cite{Ahmed:2010wy} and CDMSlite \cite{Agnese:2013jaa} experiments, with the addition of SIMPLE~\cite{Felizardo:2011uw}, PICASSO \cite{Archambault:2012pm} and COUPP \cite{Behnke:2012ys} for spin-dependent and isospin-violating interactions. CDMS-II-Ge \cite{Ahmed:2012vq} also constrains directly the amplitude of an annually modulated signal.

In order to compare a model for WIMPs with data from direct DM detection experiments, one needs to assume a value for the DM local density and velocity distribution in our galaxy. The Standard Halo Model (SHM) is usually assumed for the DM halo, corresponding to a truncated Maxwell-Boltzmann distribution for the DM velocity (see \eg \cite{Savage:2008er}). However, the parameters of this model are not known to great accuracy, and the model itself is not supported by data. Actually, quantitatively different velocity distributions are obtained from numerical simulations (see \eg \cite{Mao:2012hf}). Various models and parametrizations for the DM velocity distribution in our galaxy have been proposed as alternatives to the SHM, either derived from astrophysical data or from N-body simulations (see \eg \cite{Freese:2012xd} and references therein). Other authors have attempted to estimate the uncertainty in the determination of the properties of the DM halo, and to quantify its effects on the interpretation of DM direct detection data (see \eg \cite{Green:2002ht, McCabe:2010zh, Green:2010gw, Green:2011bv, Fairbairn:2012zs}). Another approach is that of marginalizing over the parameters of the DM halo when computing bounds and allowed regions from the experimental data (see \eg \cite{Arina:2013jma}). However, all these procedures maintain a certain degree of model dependence, \eg in the choice of the functional form of the parametrization of the halo. It is very important to notice here that the high velocity tail of the DM velocity distribution plays a crucial role in determining the number of DM particles that are above threshold for a given experiment, and therefore a way to analyze the data without the need to make any assumption on its shape is highly desirable.

The problem of comparing results from different direct detection experiments can indeed be formulated without the need to assume a velocity profile for the DM \cite{Fox:2010bz, Frandsen:2011gi, Gondolo:2012rs, Frandsen:2013cna, DelNobile:2013cta, DelNobile:2013cva, DelNobile:2013gba, DelNobile:2014eta}. The basic idea is to factor out from the formulas used to compute the scattering rate, all the astrophysical quantities such as the DM velocity distribution function. In this way the rate can be computed, for any model of particle interactions between the DM and the nuclei in the detector, with no need to assume a velocity profile for the DM, while rather allowing to use the experimental data to constrain the unknown quantities. Such a ``halo-independent'' analysis was first proposed in \cite{Fox:2010bz}, where many of the features of the method were presented, and was further developed in \cite{Frandsen:2011gi} which extended the analysis to annual modulations, and \cite{Gondolo:2012rs} which showed how to include detector resolutions. The method was further generalized in \cite{DelNobile:2013cva} to more complicate particle interactions, \ie those where the scattering cross section has a non-trivial dependence on the DM velocity.

The halo-independent analysis is particularly useful to investigate the compatibility of the different experimental results in the light WIMP hypothesis, for which the details of the DM velocity distribution, especially at high velocities, are notably relevant. Here we will review this method, applying it to spin-independent interactions with both isospin-conserving and isospin-violating \cite{Kurylov:2003ra, Feng:2011vu} couplings, and to WIMPs with magnetic dipole moment. We will compare data from DAMA \cite{Bernabei:2010mq}, CoGeNT \cite{Aalseth:2011wp}, CRESST-II \cite{Angloher:2011uu}, CDMS-II-Si \cite{Agnese:2013rvf}, CDMS-II-Ge low threshold analysis \cite{Ahmed:2010wy}, CDMS-II-Ge annual modulation analysis \cite{Ahmed:2012vq}, CDMSlite \cite{Agnese:2013jaa}, XENON10 S2-only analysis \cite{Angle:2011th}, XENON100 \cite{Aprile:2012nq}, LUX \cite{Akerib:2013tjd}, and SIMPLE \cite{Felizardo:2011uw}, following the analysis described in \cite{DelNobile:2013cta, DelNobile:2013cva, DelNobile:2013gba}. This review summarizes the results presented in \cite{DelNobile:2013cta, DelNobile:2013cva, DelNobile:2013gba}.

\section{The scattering rate}

What is observed at direct DM detection experiments is the WIMP-nucleus differential scattering rate, usually measured in units of counts/kg/day/keV. For a target nuclide $T$ initially at rest, recoiling with energy $\ER$ after the scattering with a WIMP with mass $m$ and initial velocity $\bfv$, the differential rate is
\beq
\label{dRdER}
\frac{\ud R_T}{\ud E_\text{R}} = \frac{\rho}{\mDM} \frac{C_T}{m_T} \int_{v \geqslant v_\text{min}(\ER)} \hspace{-24pt} \ud^3 v \, f(\bfv, t) \, v \, \frac{\ud \sigma_T}{\ud \ER}(\ER, \bfv) \ .
\eeq
Here $m_T$ is the target nuclide mass and $C_T$ is its mass fraction in the detector, and we denoted with $v = | \bfv |$ the WIMP speed. $\ud \sigma_T / \ud \ER$ is the differential scattering cross section. The dependence of the rate on the local characteristics of the DM halo is contained in the local DM density $\rho$ and the DM velocity distribution in the Earth's frame $f(\bfv, t)$, which is modulated in time due to Earth's rotation around the Sun \cite{Drukier:1986tm, Freese:2012xd}. The distribution $f(\bfv, t)$ is normalized to $\int \ud^3 v \, f(\bfv, t) = 1$. In the velocity integral, $\vmin(\ER)$ is the minimum speed required for the incoming DM particle to cause a nuclear recoil with energy $\ER$. For an elastic collision
\beq\label{vmin}
\vmin = \sqrt{\frac{m_T \ER}{2 \mu_T^2}} \ ,
\eeq
where $\mu_T = m \, m_T / (m + m_T)$ is the WIMP-nucleus reduced mass.

To properly reproduce the recoil rate measured by experiments, we need to take into account the characteristics of the detector. Most experiments do not measure the recoil energy directly but  rather a detected energy $\Ed$, often quoted in keVee (keV electron-equivalent) or in photoelectrons. The uncertainties and fluctuations in the detected energy corresponding to a particular recoil energy are expressed  in a (target nuclide and detector dependent) resolution function $G_T(\ER, \Ed)$, that gives the probability that a recoil energy $\ER$ (usually quoted in keVnr for nuclear recoils) is measured as $\Ed$. The resolution function is often (but not always as the XENON and LUX experiments are a notable exception) approximated by a Gaussian distribution. It incorporates the mean value $\langle \Ed \rangle = Q_T \ER$, which depends on  the energy dependent quenching factor $Q_T(\ER)$, and the energy resolution $\sigma_{\ER}(\Ed)$.  Moreover, experiments have one or more counting efficiencies or cut acceptances, denoted here as $\epsilon_1(\Ed)$ and $\epsilon_2(\ER)$, which also affect the measured rate. Thus the nuclear recoil rate in eq.~\eqref{dRdER} must be convolved with the function $\epsilon_1(\Ed) \epsilon_2(\ER) G_T(\ER, \Ed)$. The resulting differential rate as a function of the detected energy $\Ed$ is
\beq\label{dRdEd}
\frac{\ud R}{\ud \Ed} = \epsilon_1(\Ed) \sum_T \int_0^\infty \ud \ER \, \epsilon_2(\ER) G_T(\ER, \Ed) \frac{\ud R_T}{\ud \ER} \ .
\eeq
The rate within a detected energy interval $[ \Ed_1, \Ed_2]$  follows as
\begin{multline}
\label{R}
R_{[\Ed_1, \Ed_2]}(t) =  \int_{\Ed_1}^{\Ed_2} \ud\Ed \, \frac{\ud R}{\ud \Ed} =
\\
\frac{\rho}{\mDM} \sum_T \frac{C_T}{m_T} \int_0^\infty \ud \ER \, \int_{v \geqslant v_\text{min}(\ER)} \hspace{-18pt} \ud^3 v \, f(\bfv, t) \, v \, \frac{\ud \sigma_T}{\ud \ER}(\ER, \bfv)
\, \epsilon_2(\ER) \int_{\Ed_1}^{\Ed_2} \ud\Ed \, \epsilon_1(\Ed) G_T(\ER, \Ed) \ .
\end{multline}
The time dependence of the rate \eqref{R} is generally well approximated by the first terms of a harmonic series,
\beq\label{Rt}
R_{[\Ed_1, \Ed_2]}(t) = R^0_{[\Ed_1, \Ed_2]} + R^1_{[\Ed_1, \Ed_2]} \cos\!\left[ \omega (t - t_0) \right] \ ,
\eeq
where $t_0$ is the time of the maximum of the signal and $\omega = 2 \pi/$yr. The coefficients $R^0_{[\Ed_1, \Ed_2]}$ and $R^1_{[\Ed_1, \Ed_2]}$ are, respectively, the unmodulated and modulated components of the rate in the energy interval $[\Ed_1, \Ed_2]$.

\section{Halo-independent method for spin-independent interaction}
\label{haloindep-SI}

The differential cross section for the usual spin-independent (SI) interaction is
\beq\label{dsigma_T}
\frac{\ud \sigma_T}{\ud \ER} = \sigma_T^{\rm SI}(\ER) \frac{m_T}{2 \mu_T^2 v^2} \ ,
\eeq
with
\beq\label{SIcrossection}
\sigma_T^{\rm SI}(\ER) = \sigma_p \frac{\mu_T^2}{\mu_p^2} [ Z_T + (A_T - Z_T) f_n / f_p ]^2 F_{{\rm SI}, T}^2(\ER) \ .
\eeq
 Here $Z_T$ and $A_T$ are respectively the atomic and mass number of the target nuclide $T$, $F_{{\rm SI}, T}(\ER)$ is the nuclear spin-independent form factor (which we take to be the Helm form factor \cite{Helm:1956zz} normalized to $F_{{\rm SI}, T}(0) = 1$), $f_n$ and $f_p$ are the effective WIMP couplings to neutron and proton, and $\mu_p$ is the WIMP-proton reduced mass. The WIMP-proton cross section $\sigma_p$ is the parameter customarily chosen to be constrained together with the WIMP mass $m$ for SI interactions, as it does not depend on the detector and thus bounds and allowed regions from different experiments can be compared on the same plot.

The isospin-conserving coupling  $f_n = f_p$ is  usually assumed by the experimental collaborations. The isospin-violating  coupling  $f_n / f_p = -0.7$ \cite{Kurylov:2003ra, Feng:2011vu}  produces the maximum  cancellation in the expression inside the square bracket in eq.~\eqref{SIcrossection} for  xenon, thus highly suppressing the interaction cross section. This suppression is phenomenologically interesting because it weakens considerably the bounds from xenon-based detectors such as XENON and LUX which provide some of the most restrictive bounds.

Using this expression for the differential cross section, and changing integration variable from $\ER$ to $\vmin$ through eq.~\eqref{vmin}, we can rewrite eq.~\eqref{R} as
\begin{align}
\label{R1}
R^{\rm SI}_{[\Ed_1, \Ed_2]}(t) & =  \int_0^\infty \ud \vmin \, \tilde{\eta}(\vmin, t) \, \eR^{\rm SI}_{[\Ed_1, \Ed_2]}(\vmin) \ ,
\end{align}
where the velocity integral $\tilde{\eta}$ is 
\beq
\label{eta0}
\tilde{\eta}(\vmin, t) \equiv \frac{\rho \sigma_p}{\mDM} \int_{v \geqslant \vmin} \ud^3 v \, \frac{f(\bfv, t)}{v} \equiv \int_{v \geqslant \vmin} \ud^3 v \, \frac{\tilde{f}(\bfv, t)}{v} \ ,
\eeq
and we defined the response function $\eR^{\rm SI}_{[\Ed_1, \Ed_2]}(\vmin)$ for WIMPs with SI interactions as
\beq
\label{Resp_SI}
\eR^{\rm SI}_{[\Ed_1, \Ed_2]}(\vmin) \equiv
2 \vmin \sum_T \frac{C_T}{m_T} \frac{\sigma_T^{\rm SI}(\ER(\vmin))}{\sigma_p}
\, \epsilon_2(\ER(\vmin)) \int_{\Ed_1}^{\Ed_2} \ud\Ed \, \epsilon_1(\Ed) G_T(\ER(\vmin), \Ed) \ .
\eeq
Introducing the speed distribution
\begin{align}
\label{Ftilde}
\widetilde{F}(v, t) \equiv v^2 \int \ud\Omega_v \, \tilde{f}(\bfv, t) \ ,
\end{align}
we can rewrite the $\tilde{\eta}$ function as
\beq\label{etaF}
\tilde{\eta}(\vmin, t) = \int_{\vmin}^\infty \ud v \, \frac{\widetilde{F}(v, t)}{v} \ .
\eeq
The velocity integral $\tilde{\eta}(\vmin,t)$ has an annual modulation due to Earth's rotation around the Sun, and can be separated into its unmodulated and modulated components as was done for the rate in eq.~\eqref{Rt},
\beq\label{etat}
\tilde{\eta}(\vmin, t) \simeq \tilde{\eta}^0(\vmin) + \tilde{\eta}^1(\vmin) \cos\!\left[ \omega (t - t_0) \right] .
\eeq
Once the WIMP mass and interactions are fixed, the functions $\tilde{\eta}^0(\vmin)$ and $\tilde{\eta}^1(\vmin)$ are detector-independent quantities that must be common to all non-directional direct DM experiments. Thus we can map the rate measurements and bounds of different experiments  into measurements of and bounds on $\tilde{\eta}^0(\vmin)$ and $\tilde{\eta}^1(\vmin)$ as functions of $\vmin$.

\spazio

For experiments with putative DM signals, in light of eq.~\eqref{R1} we may interpret the measured rates $\hat{R}^{\, i}_{[\Ed_1, \Ed_2]} \pm \Delta{R}^{\, i}_{[\Ed_1, \Ed_2]}$ in an energy interval $[\Ed_1, \Ed_2]$ as averages of the $\tilde{\eta}^i(\vmin)$ functions weighted by the response function $\eR^{\rm SI}_{[\Ed_1, \Ed_2]}(\vmin)$:
\beq
\label{avereta}
\overline{\tilde{\eta}^{\, i}_{[\Ed_1, \Ed_2]}} \equiv \frac{\hat{R}^{\, i}_{[\Ed_1, \Ed_2]}}
{\int \ud\vmin \, \eR^{\rm SI}_{[\Ed_1, \Ed_2]}(\vmin)} \ ,
\eeq
with $i = 0, 1$ for the unmodulated and modulated component, respectively. Each such average corresponds to a point with error bars in the $(\vmin, \tilde{\eta})$ plane. The vertical bars  are given by $\Delta\overline{\tilde{\eta}^{\, i}_{[\Ed_1, \Ed_2]}}$ computed by replacing $\hat{R}^{\, i}_{[\Ed_1, \Ed_2]}$ with $\Delta{R}^{\, i}_{[\Ed_1, \Ed_2]}$  in eq.~\eqref{avereta}.  The  $\Delta{R}^{\, i}$ used here correspond to  the $68\%$ confidence interval. The horizontal bar shows the $\vmin$ interval where the response function $\eR^{\rm SI}_{[\Ed_1, \Ed_2]}(\vmin)$ for the given experiment is sufficiently different from zero. Following \cite{Frandsen:2011gi, Gondolo:2012rs, DelNobile:2013cta, DelNobile:2013gba} the horizontal bar may be chosen to extend over the interval $[{\vmin}_{,1}, {\vmin}_{,2}] = [v_{\rm min}(\Ed_1 - \sigma_{\ER}(\Ed_1)), v_{\rm min}(\Ed_2 + \sigma_{\ER}(\Ed_2))]$, where $\sigma_{\ER}(\Ed)$ is the energy resolution and the function $v_{\rm min}(\Ed)$ is obtained from $v_{\rm min}(\ER)$ in eq.~\eqref{vmin} by using the recoil energy  $\ER$ that produces the mean $\langle \Ed \rangle$ which is equal to the measured energy $\Ed$. When isotopes of the same element are present, like for Xe or Ge, the $v_{\rm min}$ intervals of the different isotopes almost completely overlap, and we take $v_{\rm min,1}$ and $v_{\rm min,2}$ to be the $C_T$-weighted averages over the isotopes of the element. When there are nuclides belonging to very different elements, like Ca and O in CRESST-II, a more complicated procedure should be followed (see \cite{Frandsen:2011gi, Gondolo:2012rs} for details).

To determine the upper bounds on the unmodulated part of $\tilde{\eta}$ set by  experimental upper bounds on the unmodulated part of the rate, a procedure first outlined in \cite{Fox:2010bz, Frandsen:2011gi} may be used.  This procedure exploits the fact that, by definition, $\tilde{\eta}^0$ is a non-increasing function of $\vmin$. For this reason, the smallest possible $\tilde{\eta}^0(\vmin)$ function passing by a fixed point $(v_0, \tilde{\eta}_0)$ in the $(\vmin, \tilde{\eta})$ plane is the downward step-function $\tilde{\eta}_0 \, \theta(v_0 - \vmin)$.  In other words, among the functions passing by the point $(v_0, \tilde{\eta}_0)$, the downward step is the function yielding the minimum predicted number of events. Imposing this functional form in eq.~\eqref{R1} we obtain
\beq
R_{[\Ed_1, \Ed_2]} = \tilde{\eta}_0 \int_0^{v_0} \ud \vmin \, \eR^{\rm SI}_{[\Ed_1, \Ed_2]}(\vmin) \ .
\eeq
The upper bound $R^{\rm lim}_{[\Ed_1, \Ed_2]}$ on the unmodulated rate in an interval $[\Ed_1, \Ed_2]$ is translated into an upper bound $\tilde{\eta}^{\rm lim}(\vmin)$ on $\tilde{\eta}^0$ at $v_0$ by
\beq
\tilde{\eta}^{\rm lim}(v_0) = \frac{R^{\rm lim}_{[\Ed_1, \Ed_2]}}{\int_0^{v_0} \ud \vmin \, \eR^{\rm SI}_{[\Ed_1, \Ed_2]}(\vmin)} \ .
\eeq
The upper bound so-obtained is conservative in the sense that any $\tilde{\eta}^0$ function extending even partially above $\tilde{\eta}^{\rm lim}$ is excluded, but not every $\tilde{\eta}^0$ function lying everywhere below $\tilde{\eta}^{\rm lim}$ is allowed \cite{Frandsen:2011gi}.

\spazio

The procedure just described does not assume any particular property of the DM halo. By making some assumptions, more stringent limits on the modulated part $\tilde{\eta}^1$ can be derived  from the limits on the unmodulated part of the rate (see \cite{Frandsen:2011gi, HerreroGarcia:2011aa, HerreroGarcia:2012fu, Bozorgnia:2013hsa}), but we choose to proceed without making any assumption on the DM halo.

\spazio

\begin{figure}[t]
\centering
\includegraphics[width=0.49\textwidth]{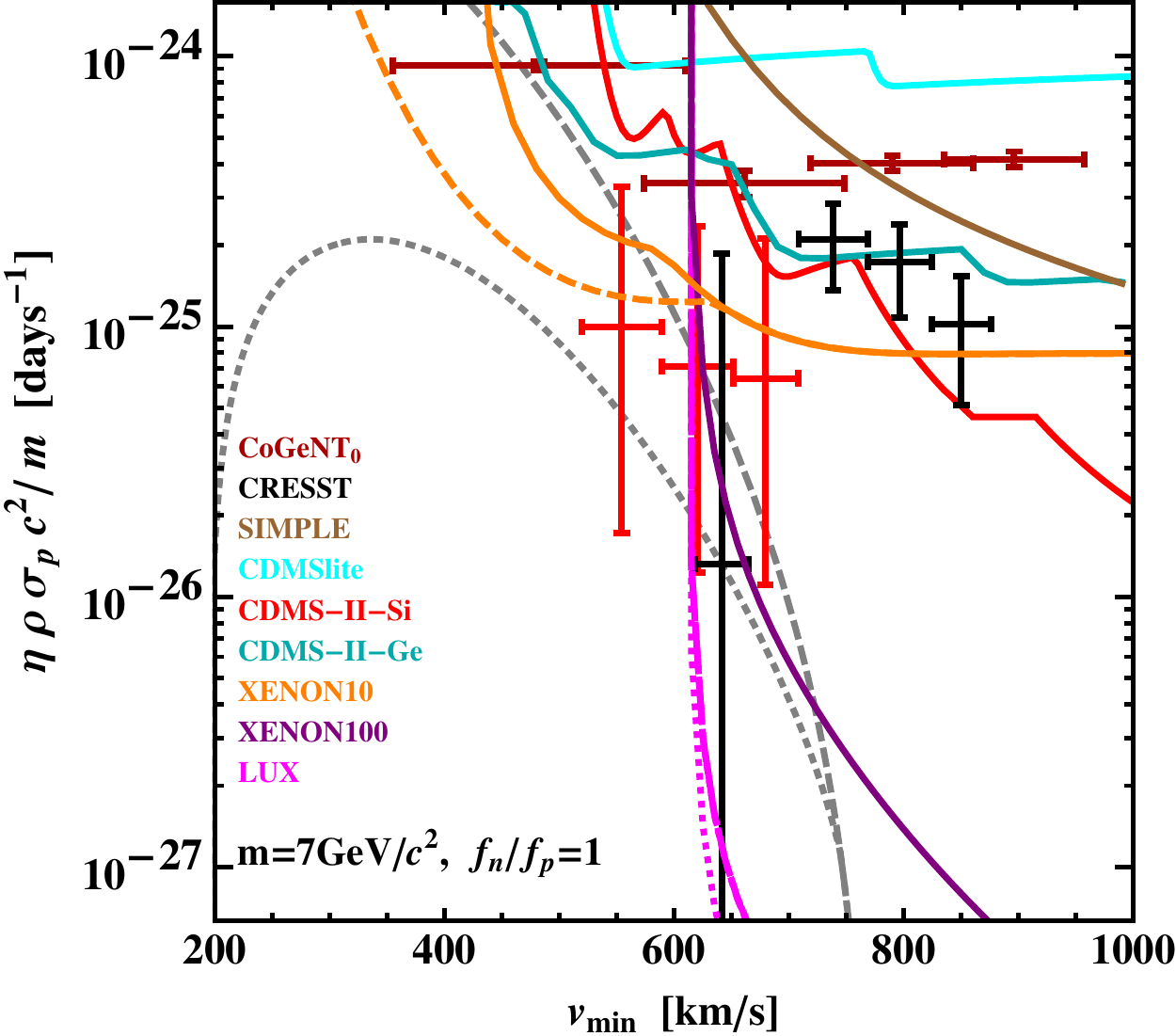}
\includegraphics[width=0.49\textwidth]{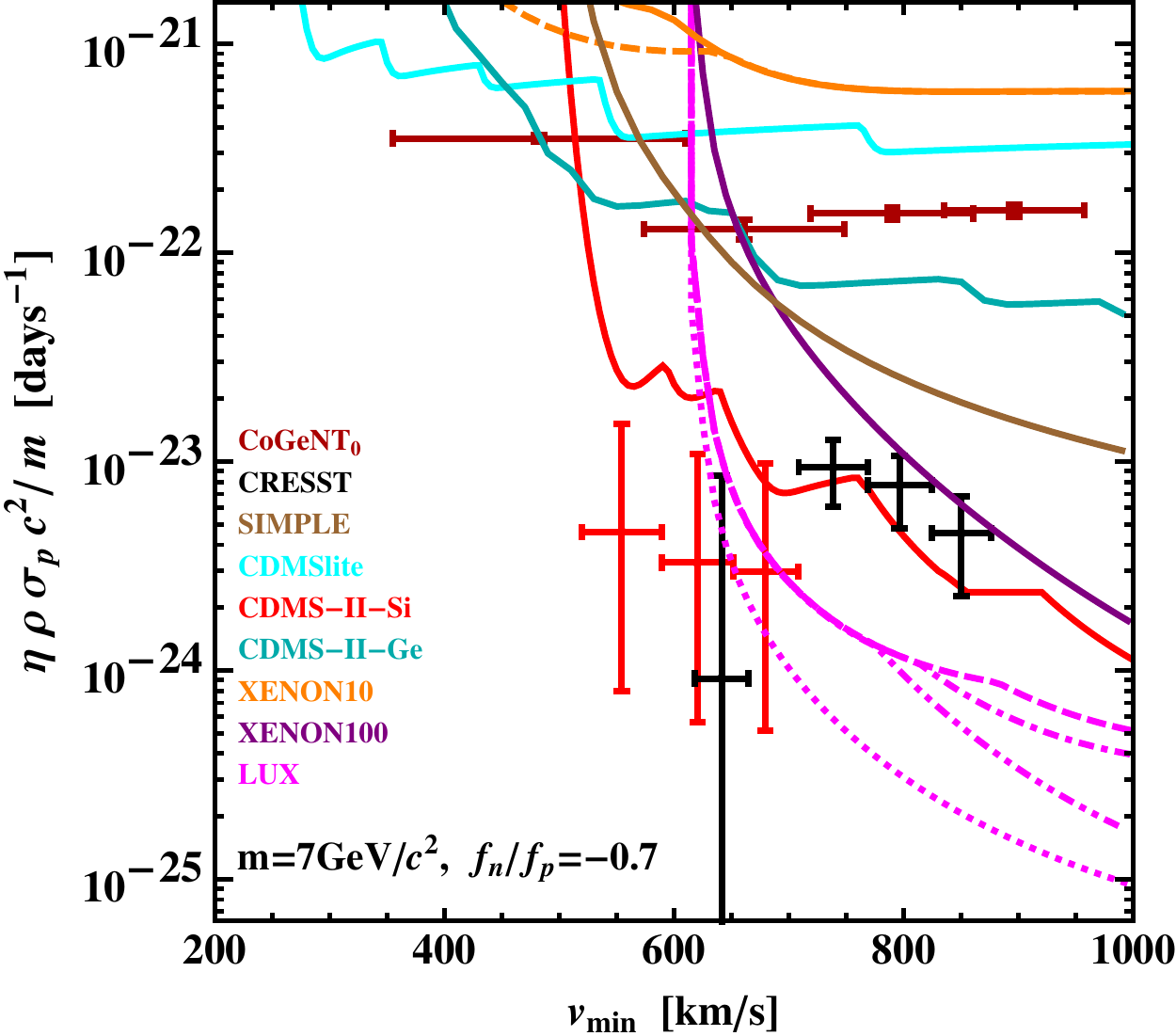}
\\
\includegraphics[width=0.49\textwidth]{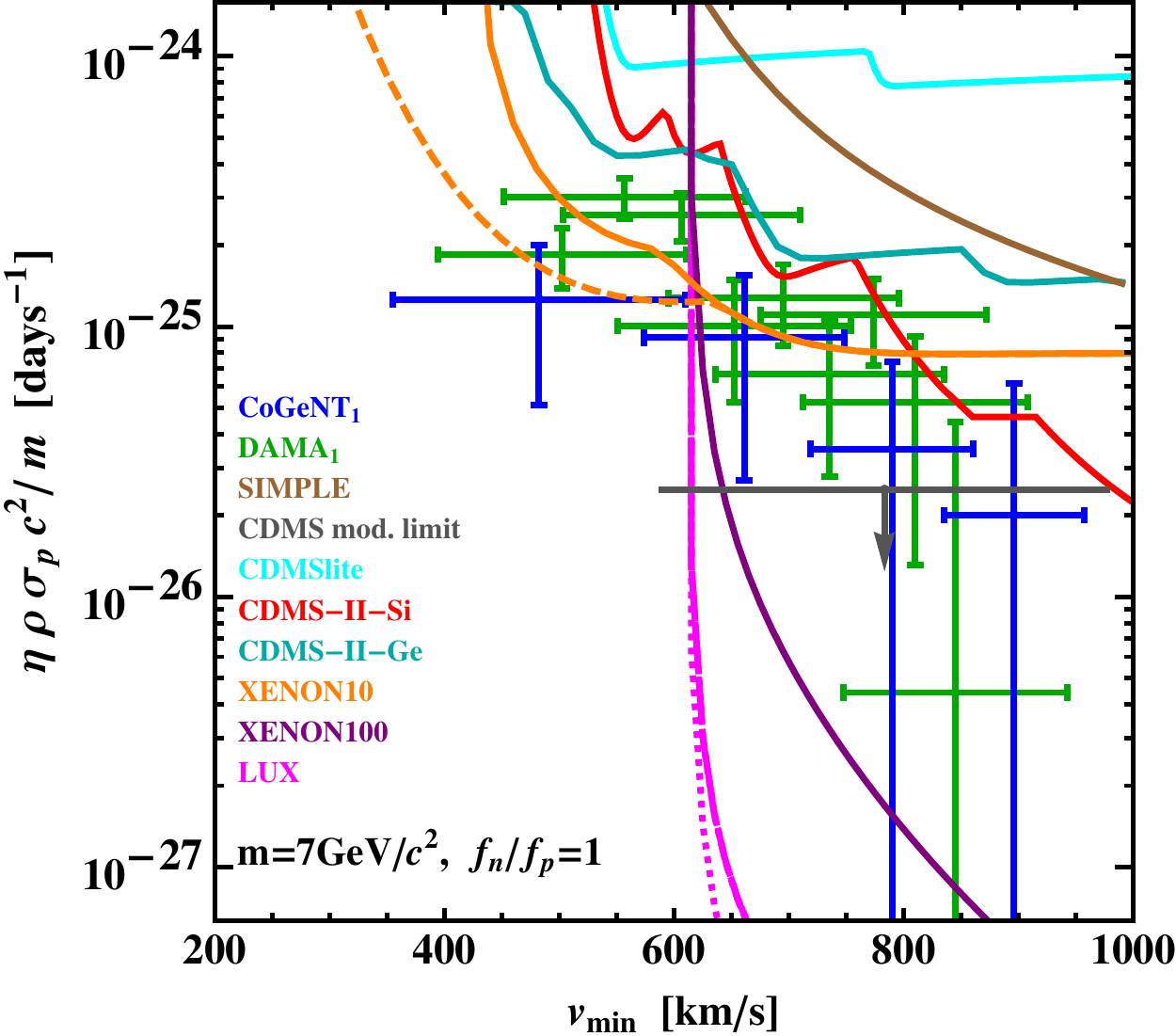}
\includegraphics[width=0.49\textwidth]{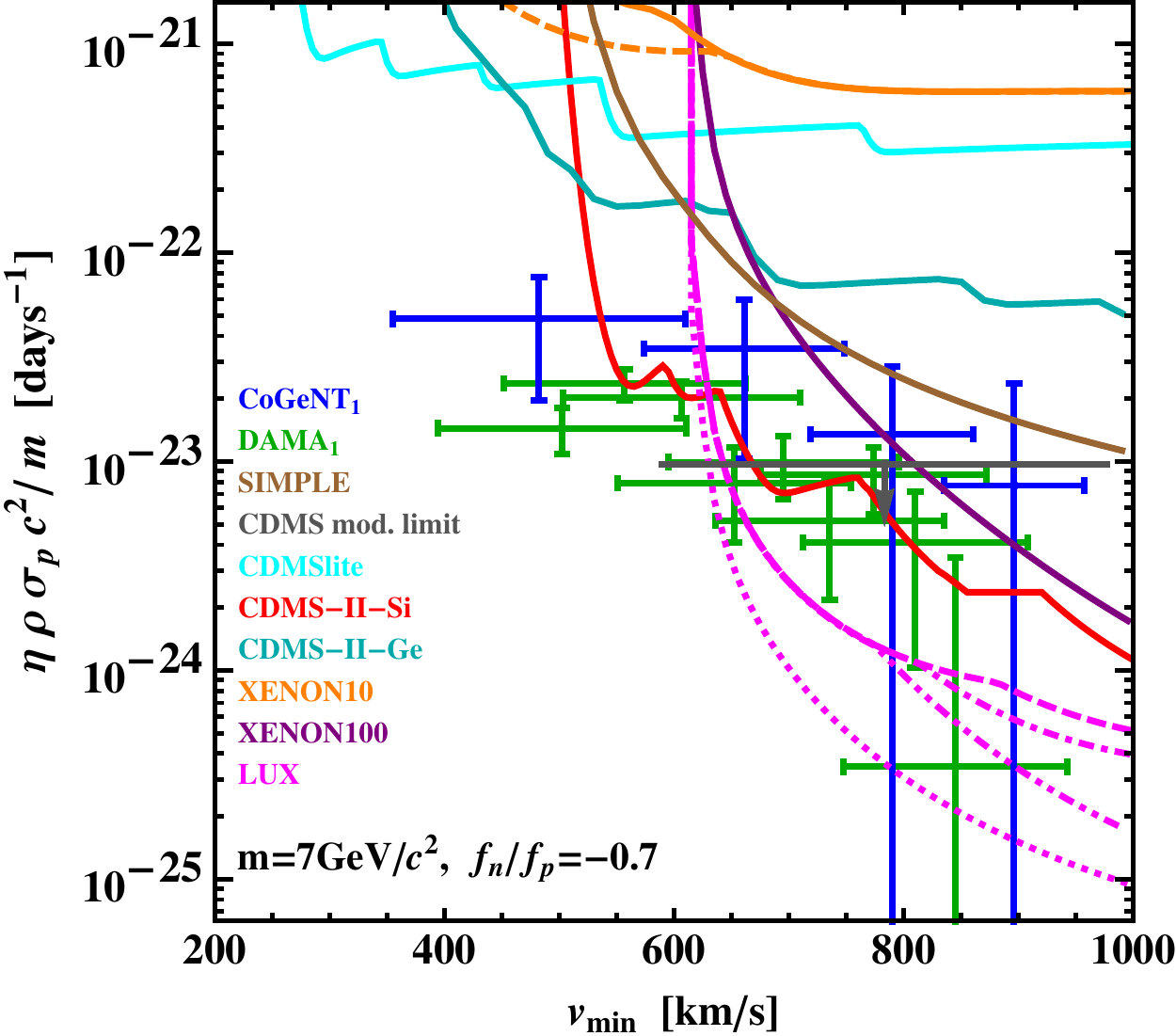}
\\
\includegraphics[width=0.49\textwidth]{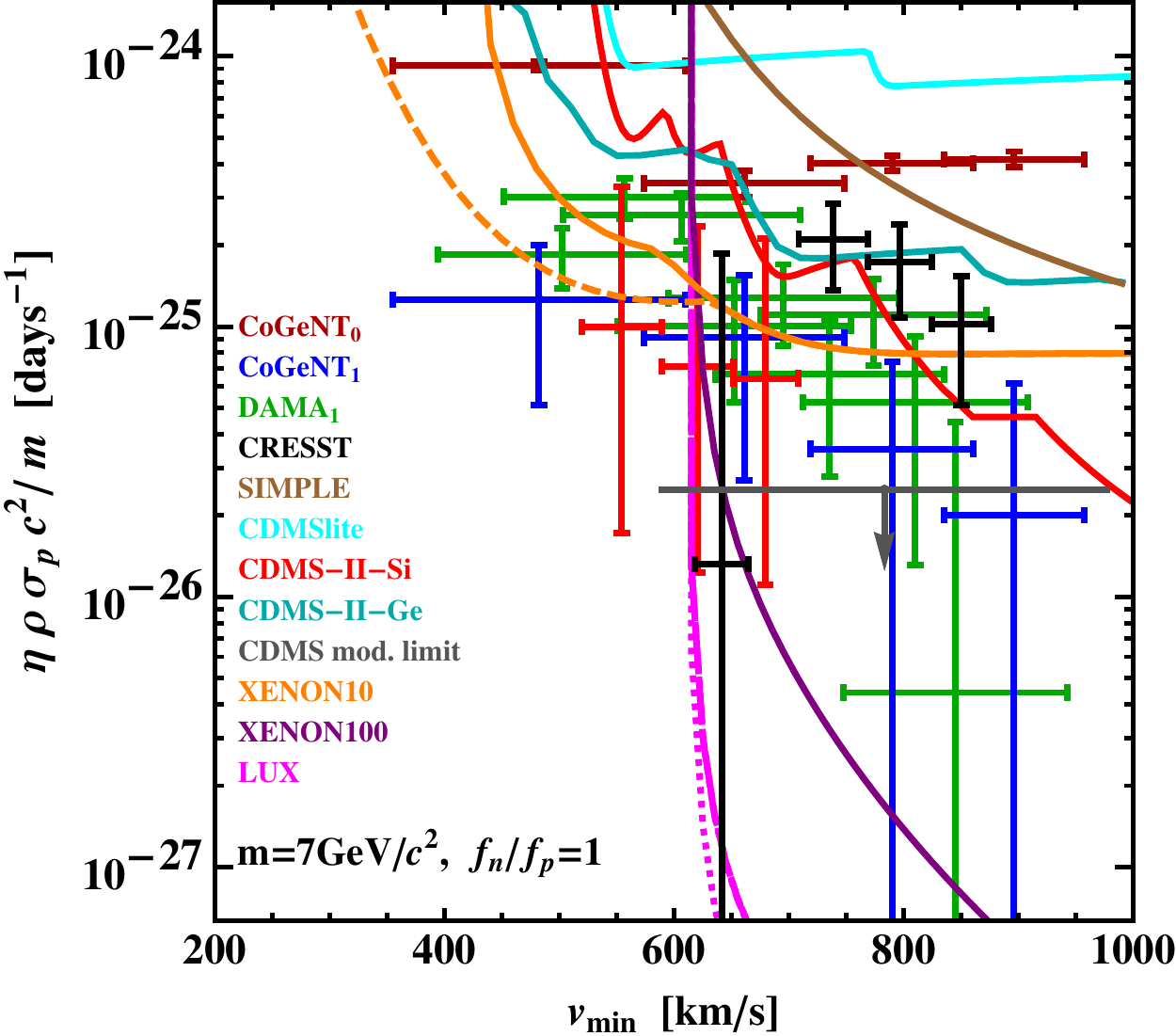}
\includegraphics[width=0.49\textwidth]{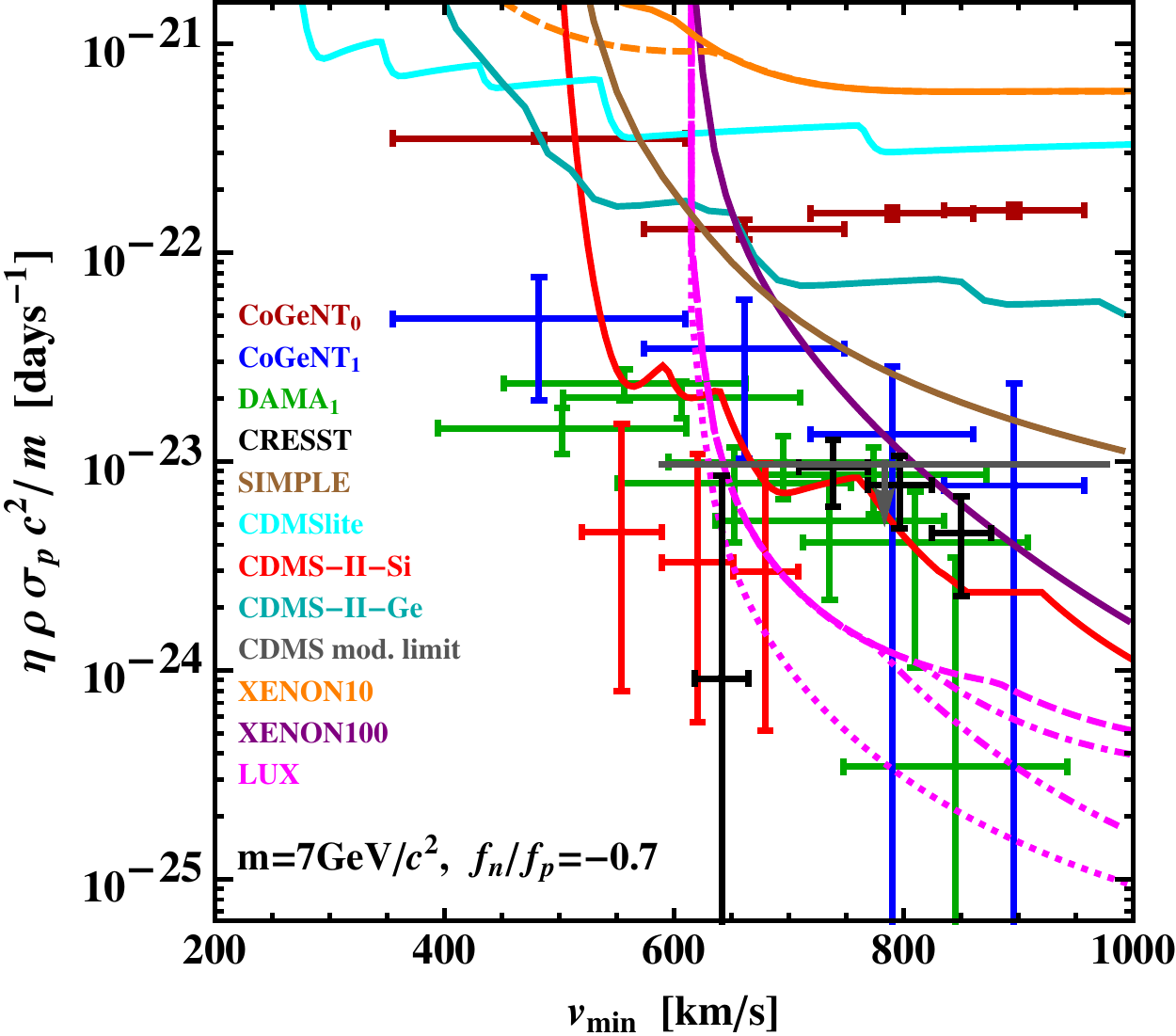}
\caption{Measurements of and bounds on $\tilde\eta^0 c^2$ and $\tilde\eta^1 c^2$ for $m = 7$ GeV/$c^2$. The left and right columns are for isospin-conserving  and isospin-violating interactions, respectively. The dashed gray lines in the top left panel show the SHM $\tilde{\eta}^0 c^2$ and $\tilde{\eta}^1 c^2$ for $\sigma_p = 10^{-40}$ cm$^2$ (see text).
}
\label{eta7}
\end{figure} 
\begin{figure}[t]
\centering
\includegraphics[width=0.49\textwidth]{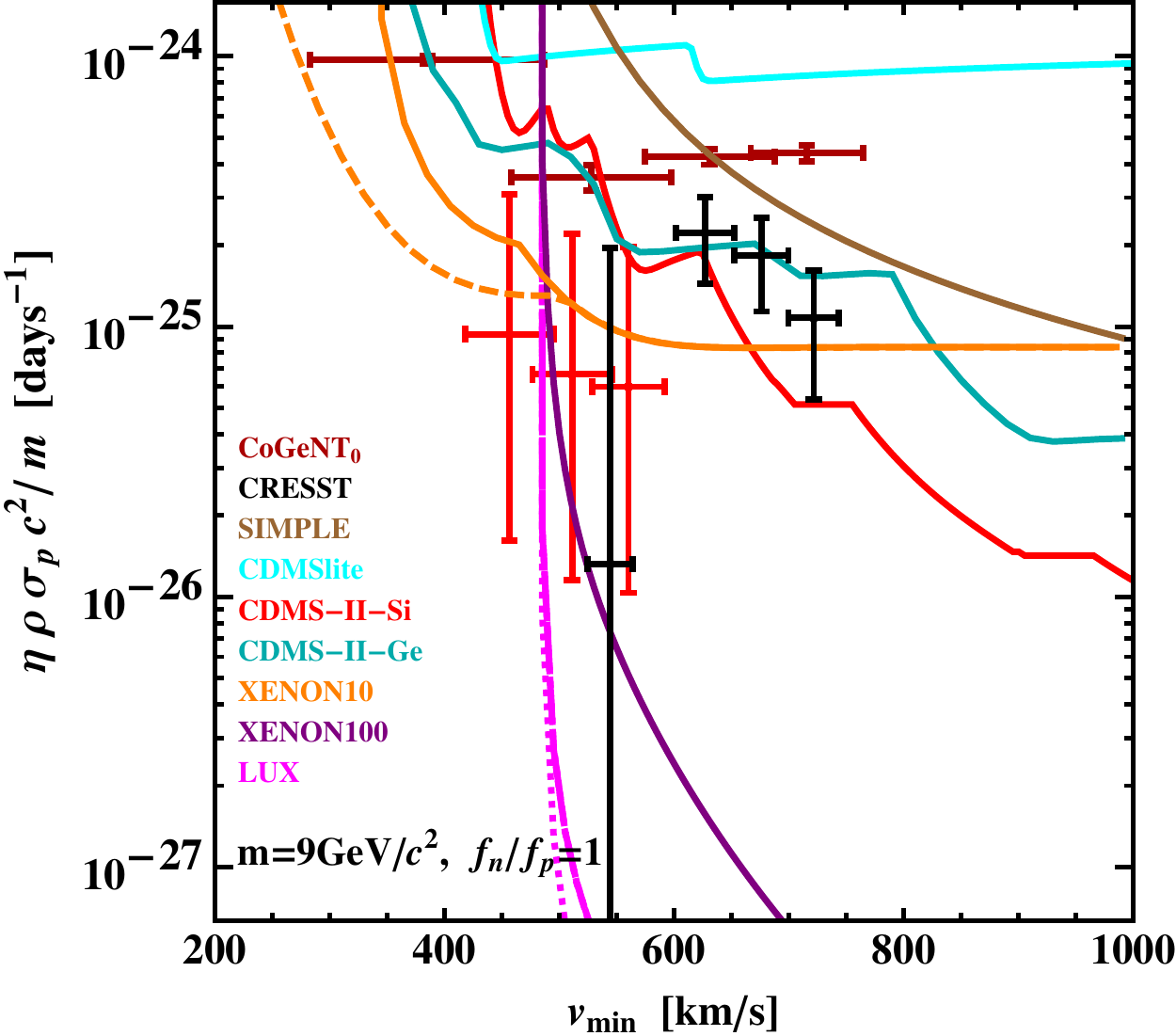}
\includegraphics[width=0.49\textwidth]{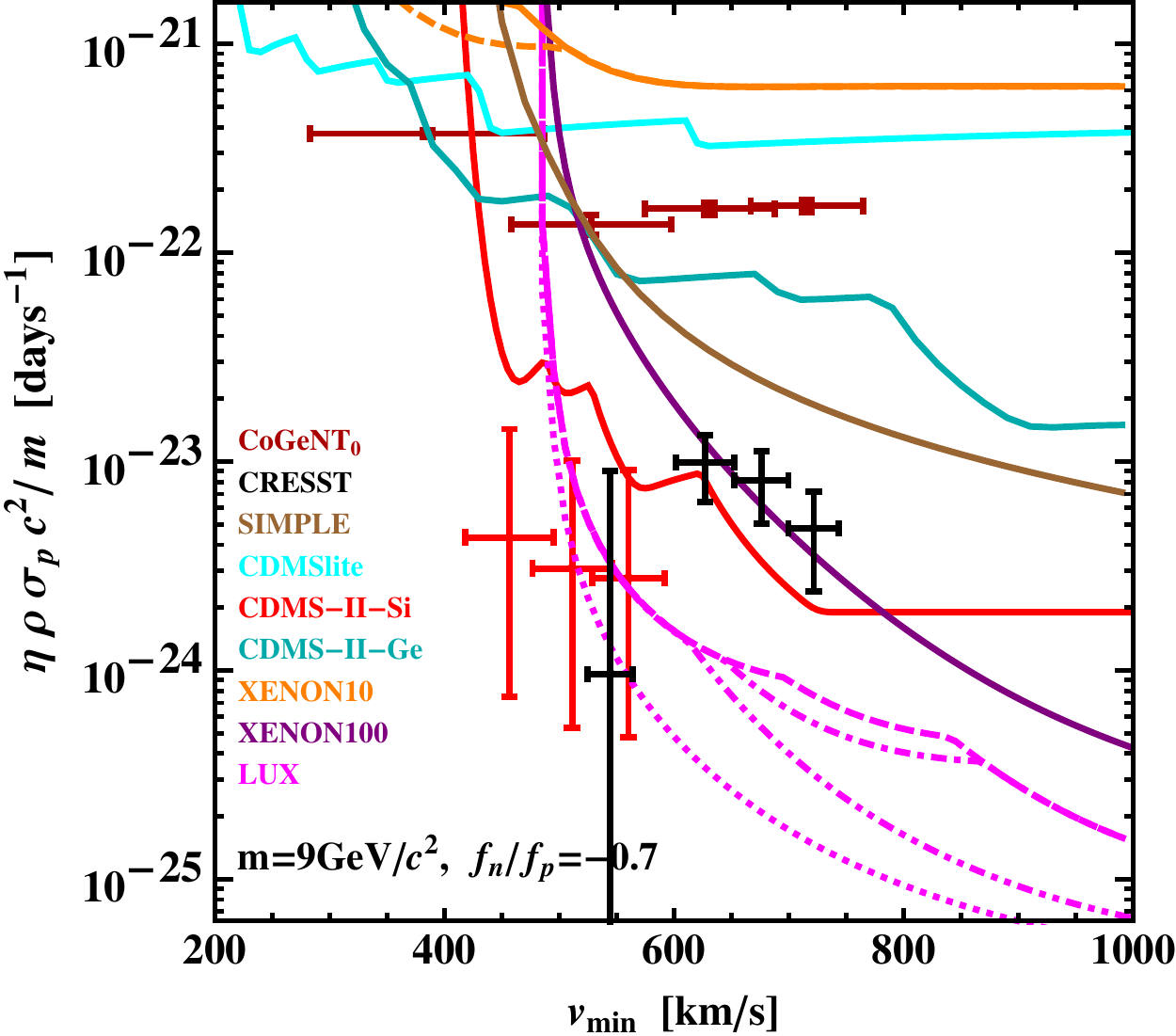}
\\
\includegraphics[width=0.49\textwidth]{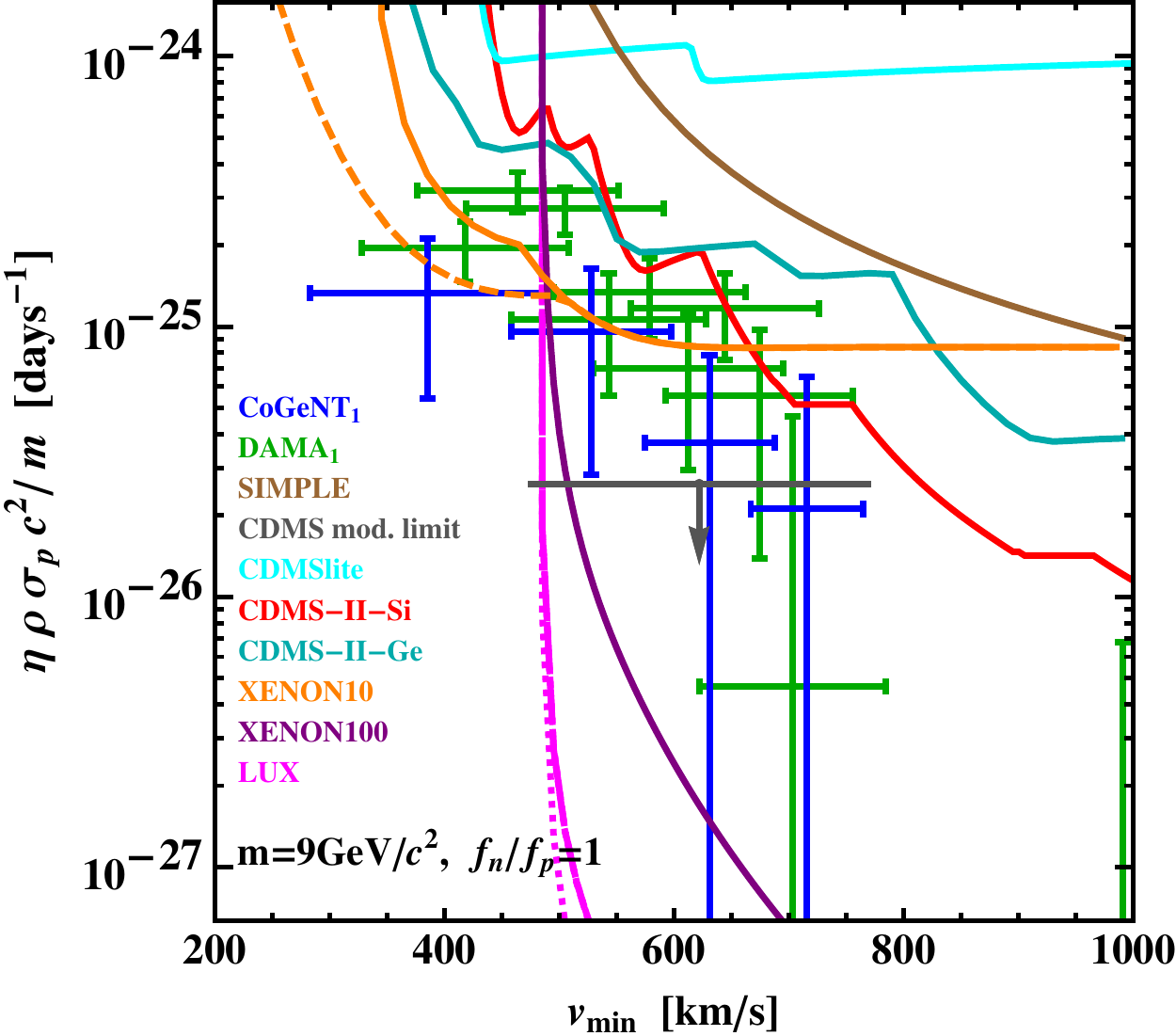}
\includegraphics[width=0.49\textwidth]{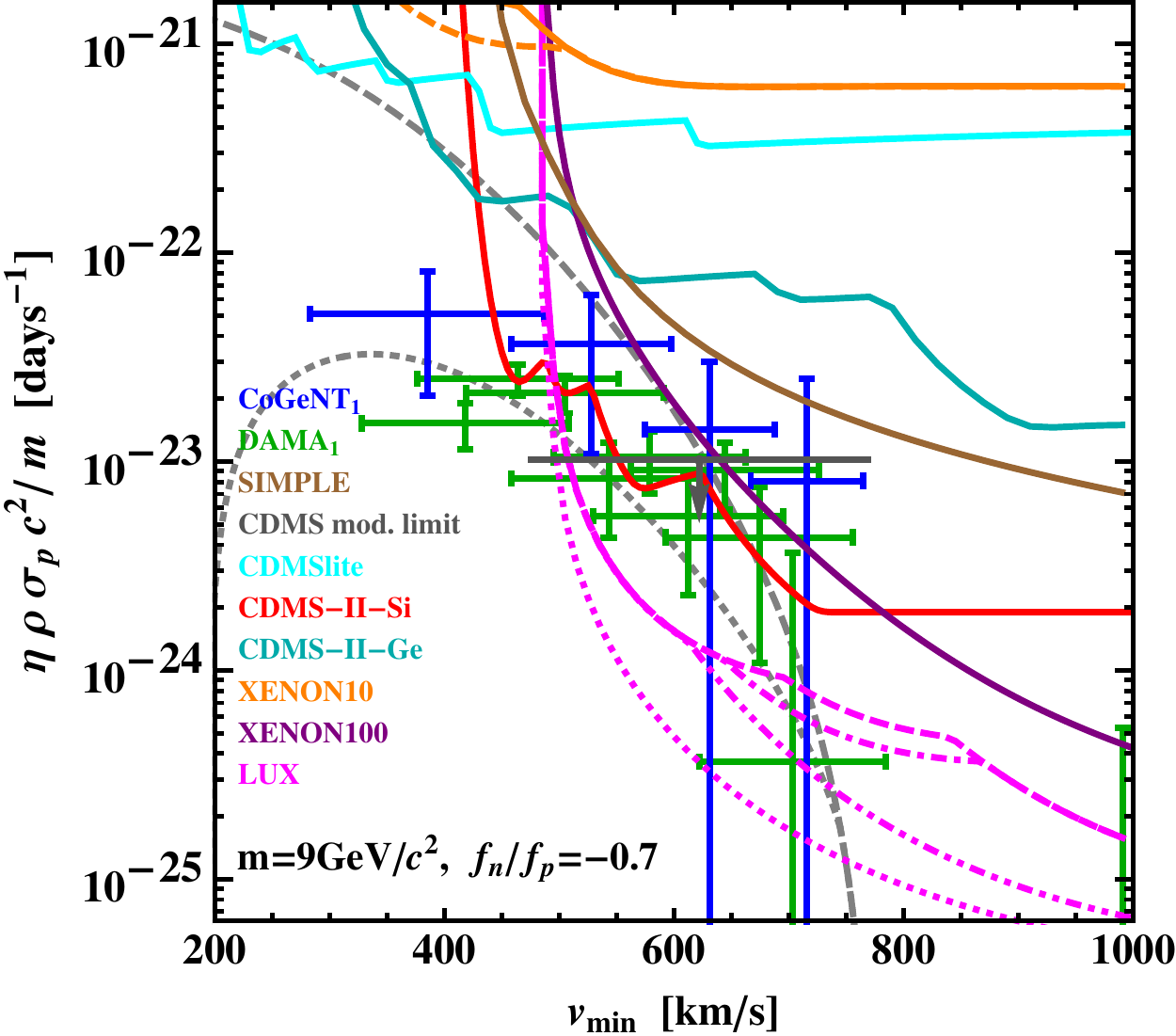}
\\
\includegraphics[width=0.49\textwidth]{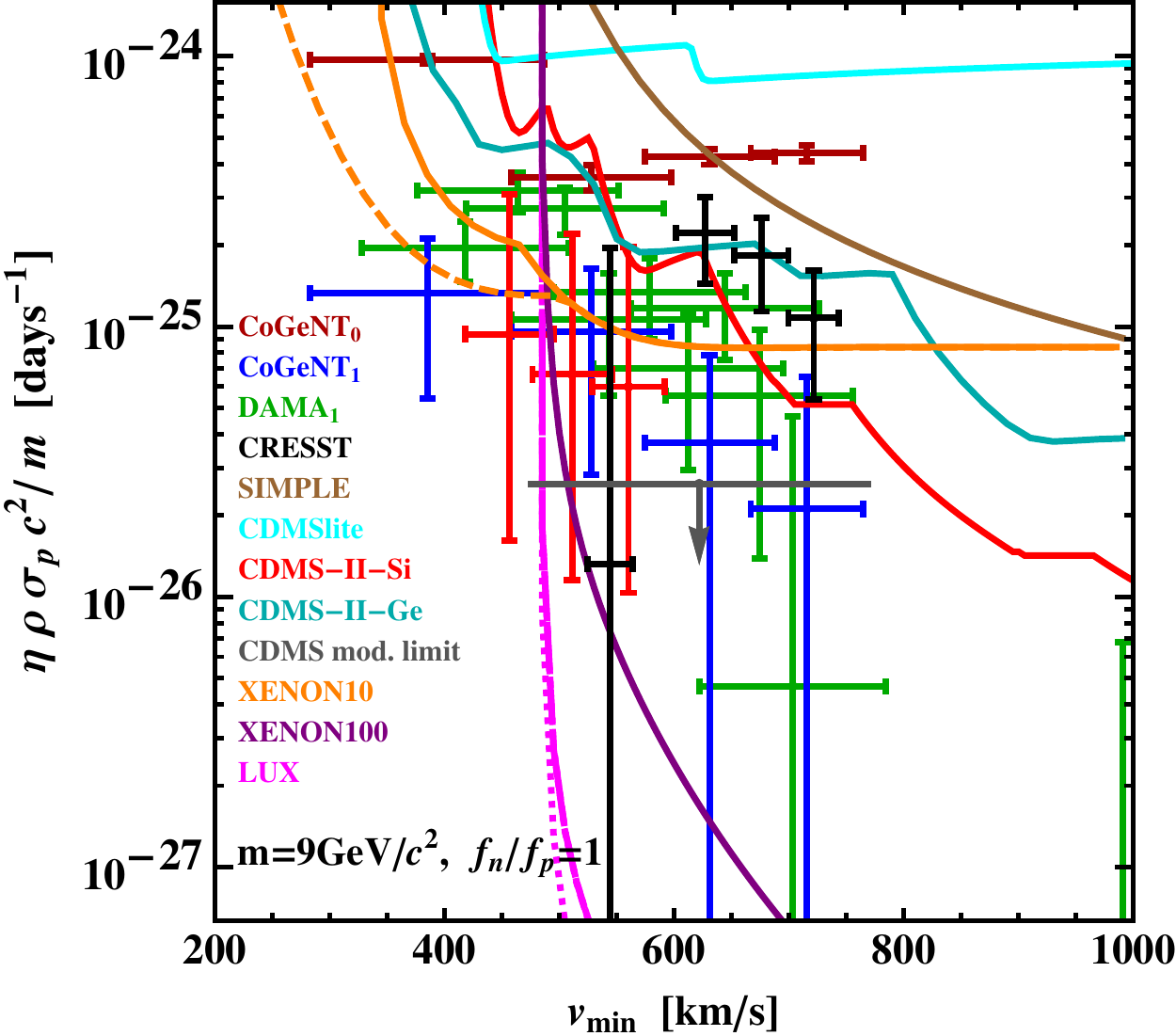}
\includegraphics[width=0.49\textwidth]{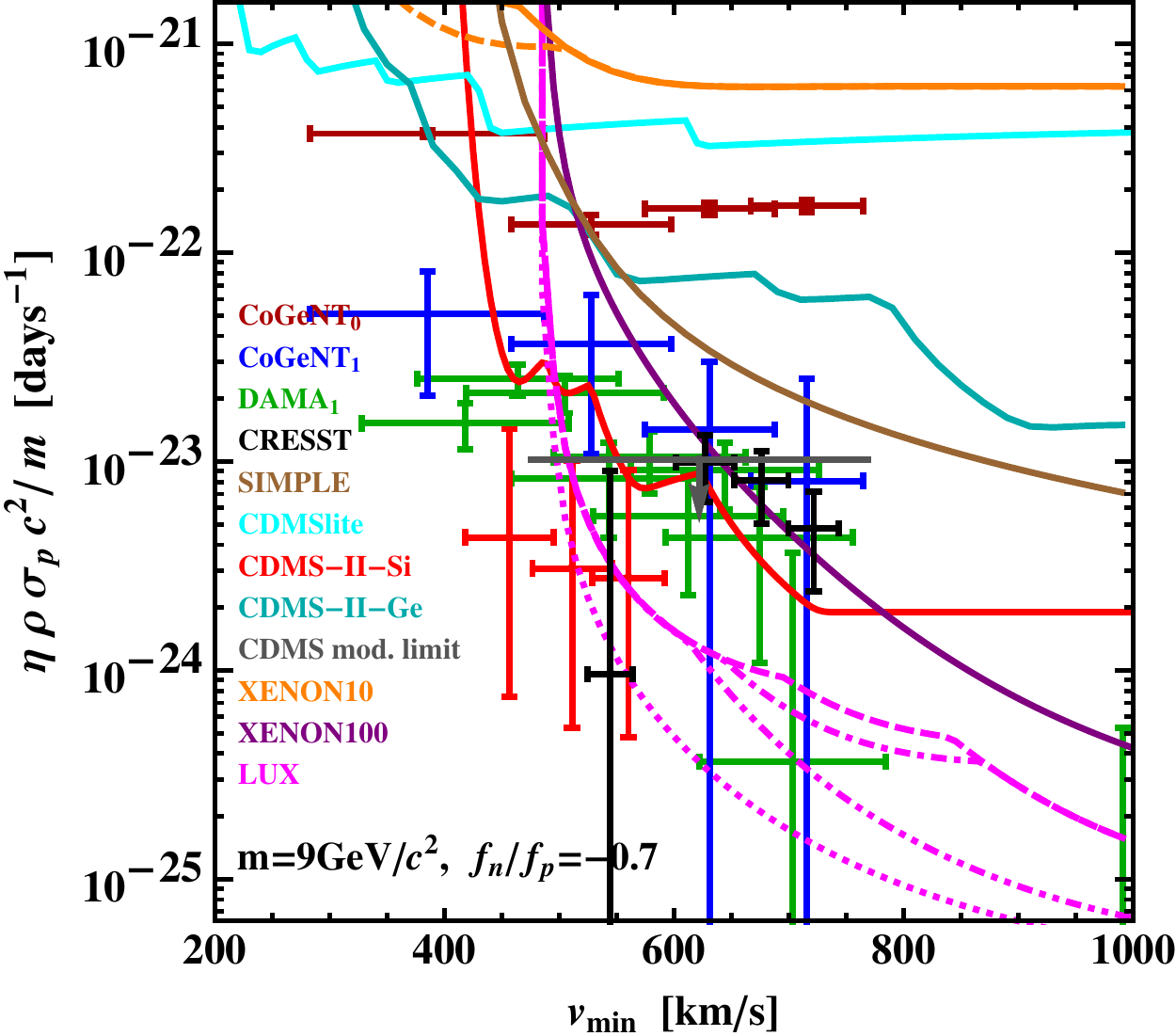}
\caption{Same as fig.~\ref{eta7}, but for a DM mass $m = 9$ GeV/$c^2$. The dashed gray lines in the middle right panel show the expected $\tilde{\eta}^0 c^2$ (upper line) and $\tilde{\eta}^1 c^2$ (lower line) for a WIMP-proton cross section $\sigma_p = 2 \times 10^{-38}$ cm$^2$ in the SHM (see text).}
\label{eta9}
\end{figure}

Figs.~\ref{eta7} and \ref{eta9} collect the results of the halo-independent analysis for a WIMP mass $m = 7$ GeV/$c^2$ and  $m = 9$ GeV/$c^2$, respectively: the left and right columns correspond to isospin-conserving ($f_n = f_p$) and isospin-violating ($f_n / f_p = -0.7$) interactions, respectively; and the top, middle and bottom rows show measurements and bounds for the unmodulated component $\tilde\eta^0 c^2$, for the modulated component $\tilde\eta^1 c^2$, and for both together, respectively, in units of day$^{-1}$. The middle row also shows the upper bounds on $\tilde\eta^0 c^2$ from the plots on the top row.

The bounds from CDMS-II-Ge, CDMS-II-Si, CDMSlite, XENON10, XENON100 and LUX are derived as $90\%$ CL upper bounds using the Maximum Gap Method \cite{Yellin:2002xd}. The SIMPLE bound is derived as the 90\% CL Poisson limit. The crosses show the DAMA modulation signal (green crosses), CoGeNT modulated (blue crosses) and unmodulated signal (plus an unknown flat background, dark red horizontal lines), CRESST-II and CDMS-II-Si unmodulated signals (black and red crosses, respectively); the CDMS-II-Ge modulation bound is shown as a dark grey horizontal line with downward arrow. Only sodium is considered for DAMA (with quenching factor $Q_{\rm Na} = 0.3$), as for the DM masses considered here the WIMP scattering off iodine is supposed to be below threshold. For XENON10, limits produced by setting or not setting the electron yield $\mathcal{Q}_{\rm y}$ to zero below 1.4 keVnr (as in \cite{Aprile:2012nq}) are obtained (solid and dashed orange line, respectively). For LUX, upper bounds considering 0, 1, 3 and 5 observed events are computed \cite{DelNobile:2013gba}, corresponding (from bottom to top) to the magenta lines with different dashing styles in figs.~\ref{eta7} and \ref{eta9}.

The overlapping of the green and blue crosses in figs.~\ref{eta7} and \ref{eta9} seem to indicate that the DAMA and CoGeNT modulation data are compatible one with the other. On the other hand, the three CDMS-II-Si points overlap or are below the CoGeNT and DAMA measurements of the modulated part of $\tilde{\eta}$. Thus, interpreted as a measurement of the unmodulated rate, the three CDMS-II-Si data points seem largely incompatible with the modulation of the signal observed by CoGeNT and DAMA, since a modulated signal is expected to be much smaller than the respective unmodulated component. For isospin-conserving interactions (left column of figs.~\ref{eta7} and \ref{eta9}), the experiments with a positive signal seem largely incompatible with the limits set by the other experiments, most notably by LUX, and by XENON10 at low $\vmin$ values. In fact, only the DAMA, CoGeNT and CDMS-II-Si data points at very low $\vmin$ are below all the bounds. The compatibility of the DAMA, CoGeNT and CRESST-II data with the exclusion bounds improves slightly for isospin-violating couplings with $f_n / f_p = -0.7$, for which the XENON and LUX limits are weakened (right column of figs.~\ref{eta7} and \ref{eta9}). However, still only the points at low $\vmin$ are below the exclusion lines, while the DAMA and CoGeNT modulated data at high $\vmin$ are now mostly excluded by the CDMS-II-Ge modulation bound.
The improvement is better for the three CDMS-II-Si data, that pass all the limits for DM with isospin-violating couplings.

In the top left panel of fig.~\ref{eta7} and the middle right panel of fig.~\ref{eta9} we show the predicted $\tilde{\eta}^0 c^2$ (upper line) and $\tilde{\eta}^1 c^2$ (lower line) from the SHM (assuming $v_0 = 220$ km/s for the DM velocity dispersion and $\vesc = 544$ km/s for the galactic escape velocity), for particular values of the WIMP-proton cross section. We choose these cross sections so that (a) for the $m = 7$ GeV/$c^2$ isospin-conserving case in the top left panel of fig.~\ref{eta7}, the CDMS-II-Si unmodulated data are well explained by the SHM, \ie the $\tilde{\eta}^0$ curve passes through the red crosses in the figure (this happens for $\sigma_p = 10^{-40}$ cm$^2$), and (b)  for the $m = 9$ GeV/$c^2$ isospin-violating case in the middle right panel of fig.~\ref{eta9},  the DAMA modulation data are well explained, \ie $\tilde{\eta}^1$ passes through the green crosses (this occurs for $\sigma_p = 2 \times 10^{-38}$ cm$^2$). These plots show that the $\tilde{\eta}^0(\vmin)$ of the SHM is a very steep function of $\vmin$ and thus can be constrained at low $\vmin$ values by the CDMSlite limit as well as by other upper limits on the unmodulated rate.

\spazio

The procedure outlined in this section to compare data from different experiments in a halo-independent way  can only be applied when the differential cross section can be factorized into a velocity dependent term, independent of the detector (\eg it must be independent of $m_T$), times a velocity independent term containing all the detector dependency. The differential cross section in eqs. \eqref{dsigma_T}, \eqref{SIcrossection} is of this form. In the case of a more general form of the differential cross section, instead, we can proceed as described in the following section.

\section{Generalized halo-independent method}\label{haloindep}

Here we show how to define the response function $\eR_{[\Ed_1, \Ed_2]}(\vmin)$ in eq.~\eqref{R1} so that the halo-independent analysis can be extended to any type of interaction. Changing the order of the $\bfv$ and $\ER$ integrations in eq.~\eqref{R}, we have
\begin{multline}\label{R2}
R_{[\Ed_1, \Ed_2]}(t) =
\\
\frac{\rho \sigma_{\rm ref}}{\mDM} \int_0^\infty \ud^3 v \, \frac{f(\bfv, t)}{v}
\sum_T \frac{C_T}{m_T} \int_0^{\ER^{\rm max}(v)} \ud \ER \, \frac{v^2}{\sigma_{\rm ref}} \frac{\ud \sigma_T}{\ud \ER}(\ER, \bfv)
\, \epsilon_2(\ER) \int_{\Ed_1}^{\Ed_2} \ud\Ed \, \epsilon_1(\Ed) G_T(\ER, \Ed) \ .
\end{multline}
Here $\ER^{\rm max}(v) \equiv 2 \mu_T^2 v^2 / m_T$ is the maximum recoil energy a WIMP of speed $v$ can impart  in an elastic collision to a target nucleus $T$ initially at rest. To make contact with the SI interaction method of the previous section, we have multiplied and divided by the factor $\sigma_{\rm ref} / v^2$, where $\sigma_{\rm ref}$ is a target-independent reference cross section (\ie a constant with the dimensions of a cross section) that coincides with $\sigma_p$  for SI interactions. In compact form, eq.~\eqref{R2} reads
\beq
R_{[\Ed_1, \Ed_2]}(t) =  \int_0^\infty \ud^3 v \, \frac{\tilde{f}(\bfv, t)}{v} \, \eH_{[\Ed_1, \Ed_2]}(\bfv) \ ,
\eeq
where in analogy with eq.~\eqref{eta0} we defined
\beq
\label{ftilde}
\tilde{f}(\bfv, t) \equiv \frac{\rho \sigma_{\rm ref}}{\mDM} \, f(\bfv, t) \ ,
\eeq
and we defined the ``integrated response function" (the name stemming from eq.~\eqref{eq:RT} below)
\beq
\label{eq:HT}
\eH_{[\Ed_1, \Ed_2]}(\bfv) \equiv
\sum_T \frac{C_T}{m_T} \int_0^{\ER^{\rm max}(v)} \ud \ER \, \frac{v^2}{\sigma_{\rm ref}} \frac{\ud \sigma_T}{\ud \ER}(\ER, \bfv)
\, \epsilon_2(\ER) \int_{\Ed_1}^{\Ed_2} \ud\Ed \, \epsilon_1(\Ed) G_T(\ER, \Ed) \  .
\eeq


 For simplicity, we only consider differential cross sections, and thus integrated response functions, that depend only on the speed $v=|\bfv|$, and not on the whole velocity vector. This is true if the DM flux and the target nuclei are unpolarized and the detection efficiency is isotropic throughout the detector, which is the most common case. With this restriction,
\begin{align}
R_{[\Ed_1, \Ed_2]}(t) & =  \int_0^\infty \ud v \, \frac{\widetilde{F}(v, t)}{v} \, \eH_{[\Ed_1, \Ed_2]}(v) \ ,
\label{eq:REEbis}
\end{align}
where the function $\widetilde{F}$ is defined as in eq.~\eqref{Ftilde}. We now define the function $\tilde{\eta}(v, t)$ by
\begin{align}
\label{eta-derivative}
\frac{\widetilde{F}(v, t)}{v}  = - \frac{ \partial \tilde{\eta}(v, t) }{\partial v} \ ,
\end{align}
with $\tilde{\eta}(v, t)$ going to zero in the limit of $v$ going to infinity. This yields the usual definition of $\tilde{\eta}$ (see eqs.~\eqref{eta0} and \eqref{etaF})
\begin{align}
\label{tildeeta}
\tilde{\eta}(v, t) = \int_v^\infty \ud v' \, \frac{\widetilde{F}(v', t)}{v'} = \int_{v' \geqslant v} \ud^3 v' \, \frac{\tilde{f}(\bfv', t)}{v'} \ .
\end{align}
Using eq.~(\ref{eta-derivative}) in eq.~(\ref{eq:REEbis})
the energy integrated rate becomes
\begin{align}
\label{R3}
R_{[\Ed_1, \Ed_2]}(t) & = - \int_0^\infty \ud v \, \frac{ \partial \tilde{\eta}(v, t) }{\partial v}  \, \eH_{[\Ed_1, \Ed_2]}(v) \ .
\end{align}
Integration by parts of eq.~(\ref{R3}) leads to an equation  formally identical to eq.~\eqref{R1} but which is now valid for any interaction,
\begin{align}
\label{R4}
R_{[\Ed_1, \Ed_2]}(t) & =  \int_0^\infty \ud \vmin \, \tilde{\eta}(\vmin, t) \,  \eR_{[\Ed_1, \Ed_2]}(\vmin) \ .
\end{align}
The response function is now defined as the derivative of the ``integrated response function" $\eH_{[\Ed_1, \Ed_2]}(v)$,
\begin{align}
\eR_{[\Ed_1, \Ed_2]}(\vmin) \equiv \left. \frac{\partial \eH_{[\Ed_1, \Ed_2]}(v)}{\partial v} \right|_{v = \vmin} .
\label{eq:RT}
\end{align}
Notice that the boundary term in the integration by parts of eq.~\eqref{R3} is zero because the definition of $\eH_{[\Ed_1, \Ed_2]}(\bfv)$ in eq.~(\ref{eq:HT}) imposes that $\eH_{[\Ed_1, \Ed_2]}(0) = 0$ (since $\ER^{\rm max}(0) = 0$).

\spazio

Similarly to what we did earlier for the SI interaction, we want again to compare average values of the $\tilde{\eta}^i$ functions with upper limits. However, for a differential cross section with a general dependence on the DM velocity, it might not be possible to simply use eq.~\eqref{avereta} with $\eR^{\rm SI}_{[\Ed_1, \Ed_2]}$ replaced by $\eR_{[\Ed_1, \Ed_2]}$ to assign a weighted  average of $\tilde{\eta}^0$ or $\tilde{\eta}^1$ to a finite $\vmin$ range. This may happen because the width of the response function $\eR_{[\Ed_1, \Ed_2]}(\vmin)$ in eq.~\eqref{eq:RT} at large $\vmin$  is dictated by the high speed behavior of the differential cross section, and it might even be infinite. For example, if $( v^2 \, \ud\sigma_T / \ud\ER)$ goes as $v^n$ for large $v$, with $n$ a positive integer, then $\eH_{[\Ed_1, \Ed_2]}(v)$ also goes as $v^n$ and  $\eR_{[\Ed_1, \Ed_2]}(\vmin)$ goes as $\vmin^{n-1}$ for large $\vmin$. Thus, if $n \geqslant 1$, the response function $\eR_{[\Ed_1, \Ed_2]}(\vmin)$ does not vanish for large $\vmin$. This implies that the denominator in eq.~\eqref{avereta} diverges.

We can regularize the behavior of the response function at large $\vmin$ by using for example the function $\vmin^r \tilde{\eta}(\vmin, t)$ with integer $r \geqslant n$, instead of just $\tilde{\eta}(\vmin, t)$.  Since this new function is common to all experiments, we can use it to compare the data in $\vmin$ space.\footnote{While any other function that goes to zero fast enough would be equally good to regularize $\eR_{[\Ed_1, \Ed_2]}(\vmin)$, as for instance an exponentially decreasing function, the power law $\vmin^{-r}$ does not require the introduction of an arbitrary $\vmin$ scale in the problem.} In fact, by multiplying and dividing the integrand in eq.~\eqref{R4} by $\vmin^r$,
\begin{align}
R_{[\Ed_1, \Ed_2]}(t) & =  \int_0^\infty \ud \vmin \, [\vmin^r \tilde{\eta}(\vmin, t)] \ [\vmin^{-r} \eR_{[\Ed_1, \Ed_2]}(\vmin)] \ ,
\end{align}
we can define the average of the functions $\vmin^r \tilde{\eta}^{\, i}(\vmin)$
\beq
\label{averetavr}
\overline{\vmin^r \tilde{\eta}^{\, i}_{[\Ed_1, \Ed_2]}} \equiv \frac{\hat{R}^{\, i}_{[\Ed_1, \Ed_2]}}{\int^\infty_0 \ud \vmin \, \vmin^{-r} \, \eR_{[\Ed_1, \Ed_2]}(\vmin)}
\eeq
($i=0$ for the unmodulated and $i=1$ for the modulated component, see eq.~\eqref{etat}). Notice that exploiting the definition of $\eR_{[\Ed_1, \Ed_2]}$ in eq.~\eqref{eq:RT}, we can write this relation in terms of $\eH_{[\Ed_1, \Ed_2]}$ instead of $\eR_{[\Ed_1, \Ed_2]}$ as
\beq
\label{averetavr2}
\overline{\vmin^r \tilde{\eta}^{\, i}_{[\Ed_1, \Ed_2]}} \equiv \frac{\hat{R}^{\, i}_{[\Ed_1, \Ed_2]}}{r \int^\infty_0 \ud v \, v^{-r-1} \, \eH_{[\Ed_1, \Ed_2]}(v)} \ ,
\eeq
where in the integration by parts the  finite term $\left[ v^{-r} \, \eH_{[\Ed_1, \Ed_2]}(v) \right]^\infty_0$ vanishes since by assumption $r$ has been appropriately chosen to regularize the integral of $\vmin^{-r} \eR_{[\Ed_1, \Ed_2]}(\vmin)$, \ie $v^{-r} \, \eH_{[\Ed_1, \Ed_2]}(v)\to 0$ as $v \to \infty$.

Eqs.~\eqref{averetavr} or \eqref{averetavr2} allow to translate rate measurements in a detected energy interval $[\Ed_1, \Ed_2]$ into averaged values of $\vmin^r \tilde{\eta}^{\, i}(\vmin)$ in a finite $\vmin$ interval $[{\vmin}_{,1}, {\vmin}_{,2}]$. This is now the interval outside which the integral of  the new response function $\vmin^{-r} \, \eR_{[\Ed_1, \Ed_2]}(\vmin)$ (and not of $\eR_{[\Ed_1, \Ed_2]}(\vmin)$) is negligible. We choose to use $90\%$ central quantile intervals, \ie  we determine ${\vmin}_{,1}$ and ${\vmin}_{,2}$ such that the area under the function  $\vmin^{-r} \, \eR_{[\Ed_1, \Ed_2]}(\vmin)$ to the left of ${\vmin}_{,1}$ is $5\%$ of the total area, and the area to the right of ${\vmin}_{,2}$ is also $5\%$ of the total area. In practice, the larger the value of $r$, the smaller is the width of the $[{\vmin}_{,1}, {\vmin}_{,2}]$  interval, designated by  the horizontal error bar of the crosses in the $(\vmin, \tilde{\eta})$ plane. However, $r$  cannot be chosen arbitrarily large, because  large values of $r$ give a large weight to the low velocity tail of the $\eR_{[\Ed_1, \Ed_2]}(\vmin)$ function, and this tail depends on the low energy tail of the resolution function $G_T(\ER, \Ed)$ in eq.~\eqref{eq:HT}, which is never well known. Therefore too large values of $r$ make the procedure very sensitive to the way in which the tails of the $G_T(\ER, \Ed)$ function are modeled. This is shown more explicitly in sec.~\ref{sec:magneticDM} (see also fig.~\ref{fig:responsefunctioncogent}), where we apply this procedure to a particular WIMP-nuclei interaction. 
In the figures, the  horizontal placement of the vertical bar in the crosses corresponds to the maximum of $\vmin^{-r} \, \eR_{[\Ed_1, \Ed_2]}(\vmin)$. The extension of the vertical bar, unless otherwise indicated, shows the 1$\sigma$ interval around the central value of the measured rate.

The upper limit on the unmodulated part of $\vmin^r \tilde{\eta}$ is simply $\vmin^r \tilde{\eta}^{\rm lim}(\vmin)$, where $\tilde{\eta}^{\rm lim}(\vmin)$ is computed as described in sec.~\ref{haloindep-SI} by using a downward step-function $\tilde{\eta}_0 \, \theta(v_0 - \vmin)$ for $\tilde{\eta}^0(v_0)$ to determine the maximum value of the step $\tilde{\eta}_0$. Given the definition of the response function $\eR$ in the general case in terms of $\eH$, eq.~\eqref{eq:RT}, the downward step-function choice for  $\tilde{\eta}^0$ yields
\beq
R_{[\Ed_1, \Ed_2]} = \tilde{\eta}_0 \int_0^{v_0} \ud \vmin \, \eR_{[\Ed_1, \Ed_2]}(\vmin) = \tilde{\eta}_0 \, \eH_{[\Ed_1, \Ed_2]}(v_0) \ .
\eeq
From this equation we find the maximum value of $\tilde{\eta}_0$ at $v_0$ allowed by the experimental upper limit on the unmodulated rate $R^{\rm lim}_{[\Ed_1, \Ed_2]}$,
\beq
\tilde{\eta}^{\rm lim}(v_0) = \frac{R^{\rm lim}_{[\Ed_1, \Ed_2]}}{\int_0^{v_0} \ud \vmin \, \eR_{[\Ed_1, \Ed_2]}(\vmin)}
= \frac{R^{\rm lim}_{[\Ed_1, \Ed_2]}}{\eH_{[\Ed_1, \Ed_2]}(v_0)} \ .
\eeq
In the figures, rather than drawing the new averages $\overline{\vmin^r \tilde{\eta}^{\, i}} c^2$ and the limits $\vmin^r \tilde{\eta}^{\rm lim}(\vmin) c^2$, we may draw $\vmin^{-r} \overline{\vmin^r \tilde{\eta}^{\, i}} c^2$ and $\tilde{\eta}^{\rm lim}(\vmin) c^2$ (in units of day$^{-1}$), so that a comparison can be easily made with the results obtained for the SI interaction shown in the previous section.

\section{Application to magnetic dipole dark matter}
\label{sec:magneticDM}

We apply here the generalized halo-independent method to a Dirac fermion DM candidate that interacts only through an anomalous magnetic dipole moment $\lambda_\chi$ (see \eg \cite{Pospelov:2000bq, An:2010kc, Sigurdson:2004zp, Barger:2010gv, Chang:2010en, Cho:2010br, Heo:2009vt, Gardner:2008yn, Masso:2009mu, Banks:2010eh, Fortin:2011hv, Kumar:2011iy, Barger:2012pf, DelNobile:2012tx, Cline:2012is, Weiner:2012cb, Tulin:2012uq, Cline:2012bz}),
\beq
\label{magneticDMlagrangian}
\Lag_{\rm int} = \frac{\lambda_\chi}{2} \, \bar\chi \sigma_{\mu \nu} \chi \, F^{\mu\nu} \ .
\eeq
The differential cross section for scattering of a magnetic dipole dark matter (MDM) with a target nucleus is
\beq
\label{magneticDMsigma}
\frac{\ud \sigma_T}{\ud \ER} =
\alpha \lambda_\chi^2 \left\{ Z_T^2 \left[\frac{1}{\ER} - \frac{1}{v^2} \frac{m_T}{2} \left( \frac{1}{\mu_T^2} - \frac{1}{\mDM^2} \right) \right] F_{{\rm E}, T}^2(\ER)
+ \frac{\hat\lambda_T^2}{v^2} \frac{m_T}{m_p^2} \left( \frac{S_T+1}{3S_T} \right) F_{{\rm M}, T}^2(\ER) \right\} .
\eeq
 Here $\alpha = e^2 / 4 \pi$ is the electromagnetic fine structure constant, $m_p$ is the proton mass, $S_T$ is the spin of the target nucleus, and $\hat\lambda_T$ is the magnetic moment of the target nucleus in units of the nuclear magneton $ e / (2 m_p)= 0.16$ (GeV/$c^2$)$^{-1}$. The first term corresponds to the dipole-nuclear charge coupling, and $F_{{\rm E}, T}(\ER)$ is the corresponding nuclear charge form factor. We take it to be the Helm form factor \cite{Helm:1956zz} normalized to $F_{{\rm E}, T}(0) = 1$. The second  term, which we call ``magnetic", corresponds to the coupling of the DM magnetic dipole to the magnetic field of the nucleus, and the corresponding nuclear form factor is the nuclear magnetic form factor $F_{{\rm M}, T}(\ER)$. This magnetic form factor is not identical to the spin form factor that accompanies spin-dependent interactions, in that the magnetic form factor includes the magnetic currents due to the orbital motion of the nucleons in addition to the intrinsic nucleon magnetic moments (spins). For the light WIMPs considered here, the magnetic term is negligible for all the target nuclei we consider except Na. This term is more important for lighter nuclei, such as Na and Si, but only $^{23}$Na has a non-negligible magnetic dipole moment, $\hat\lambda_{\rm Na}=$ 2.218. The magnetic form factor for this nuclide is taken from \cite{Donnelly:1984rg} as explained in \cite{DelNobile:2013cva}.

The spin-independent part of the differential cross section has two terms, with different $v$ dependences. Therefore, had we computed the rate with the method used to get to eq.~\eqref{R1}, we would have obtained two terms  in the rate each containing a different function of $\vmin$ multiplied by detector dependent coefficients. It would have been impossible in this way to translate a rate measurement or bound into only one of the two $\vmin$ functions. In such a situation, the approach presented in sec.~\ref{haloindep-SI} can not be applied and one needs to resort to the generalized method of sec.~\ref{haloindep}. The function $\eH_{[\Ed_1, \Ed_2]}(v)$ has in this case a $v^2$ dependence for large values of $v$, with $\eR_{[\Ed_1, \Ed_2]}(\vmin)$ scaling as $\vmin$. More precisely we have

\begin{multline}
\eH_{[\Ed_1, \Ed_2]}(\bfv) =
\sum_T \frac{C_T}{m_T} \int_0^{\ER^{\rm max}(v)} \ud \ER
\\
\times
\left\{ Z_T^2 \left[\frac{v^2}{\ER} - \frac{m_T}{2} \left( \frac{1}{\mu_T^2} - \frac{1}{\mDM^2} \right) \right] F_{{\rm E}, T}^2(\ER) + \hat\lambda_T^2 \frac{m_T}{m_p^2} \left( \frac{S_T+1}{3S_T} \right) F_{{\rm M}, T}^2(\ER) \right\}
\\
\times \epsilon_2(\ER) \int_{\Ed_1}^{\Ed_2} \ud\Ed \, \epsilon_1(\Ed) G_T(\ER, \Ed) \ ,
\end{multline}
where we defined $\sigma_{\rm ref} \equiv \alpha \lambda_\chi^2$. As a consequence,
\begin{multline}
\label{R_MDM}
\eR_{[\Ed_1, \Ed_2]}(\vmin) =
2 \vmin \sum_T \frac{C_T}{m_T} \int_0^\infty \ud \ER
\\
\times
\left\{ \left[ Z_T^2 \frac{\mu_T^2}{\mDM^2} F_{{\rm E}, T}^2(\ER) + \hat\lambda_T^2 \frac{2 \mu_T^2}{m_p^2} \left( \frac{S_T+1}{3S_T} \right) F_{{\rm M}, T}^2(\ER) \right] \delta(\ER^{\rm max}(\vmin) - \ER) \right.
\\
\left. + Z_T^2 \frac{1}{\ER} F_{{\rm E}, T}^2(\ER) \, \theta(\ER^{\rm max}(\vmin) - \ER) \right\}
\epsilon_2(\ER) \int_{\Ed_1}^{\Ed_2} \ud\Ed \, \epsilon_1(\Ed) G_T(\ER, \Ed) \ .
\end{multline}
The denominator of eq.~\eqref{averetavr} is therefore
\begin{multline}
\int \ud \vmin \, \vmin^{-r} \, \eR_{[\Ed_1, \Ed_2]}(\vmin) =
2 \sum_T \frac{C_T}{m_T} \int_0^\infty \ud \vmin \, \vmin^{-r + 1}
\\
\times
\left\{ Z_T^2 \left( \frac{\mu_T^2}{\mDM^2} + \frac{2}{r-2} \right) F_{{\rm E}, T}^2(\ER(\vmin)) + \hat\lambda_T^2 \frac{2 \mu_T^2}{m_p^2} \left( \frac{S_T+1}{3S_T} \right) F_{{\rm M}, T}^2(\ER(\vmin)) \right\}
\\
\times \epsilon_2(\ER) \int_{\Ed_1}^{\Ed_2} \ud\Ed \, \epsilon_1(\Ed) G_T(\ER(\vmin), \Ed) \ ,
\end{multline}
where $r$ can be any number larger than $2$. To obtain this result, we first integrated the $\theta$ term in eq.~\eqref{R_MDM} with respect to $\vmin$, and then used eq.~\eqref{vmin} to change integration variable from $\ER$ to $\vmin$ again.

\begin{figure}[t]
\centering
\includegraphics[width=0.49\textwidth]{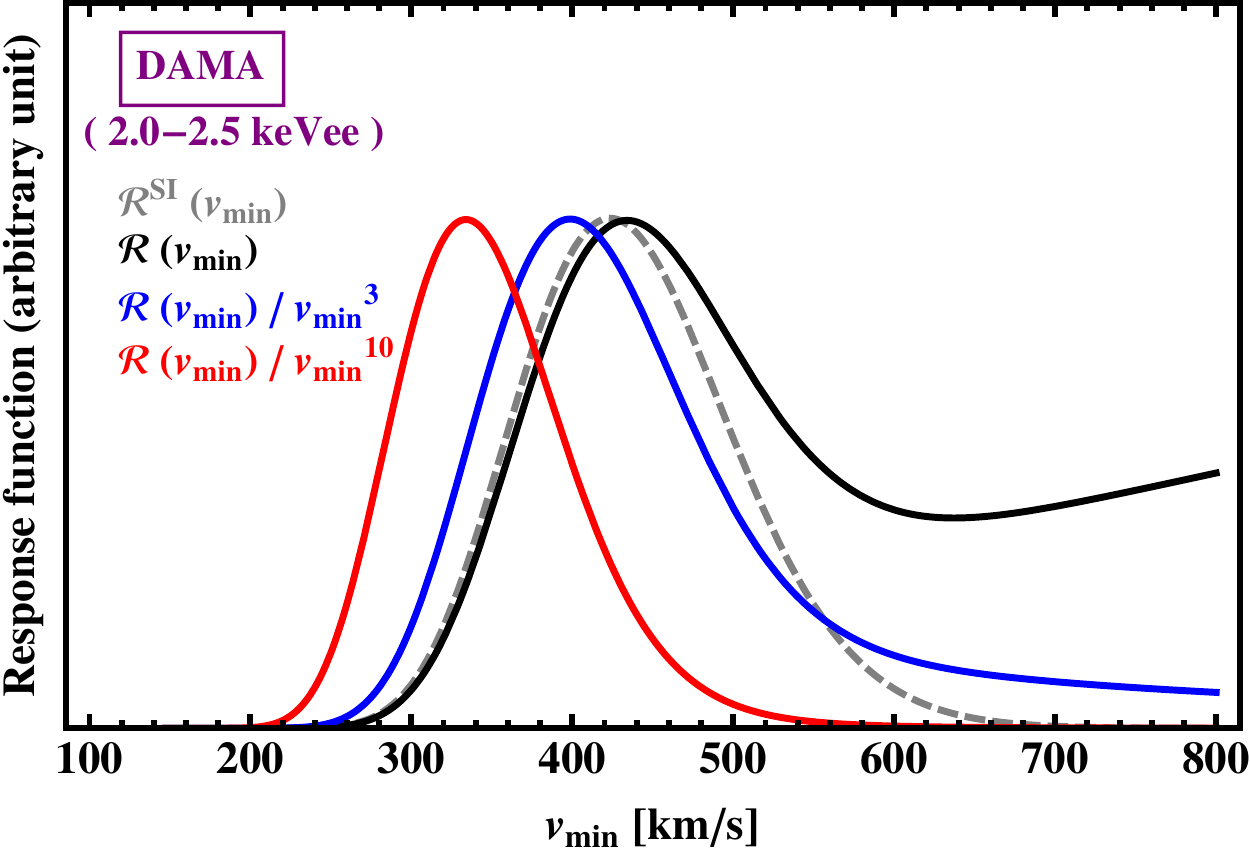}
\includegraphics[width=0.49\textwidth]{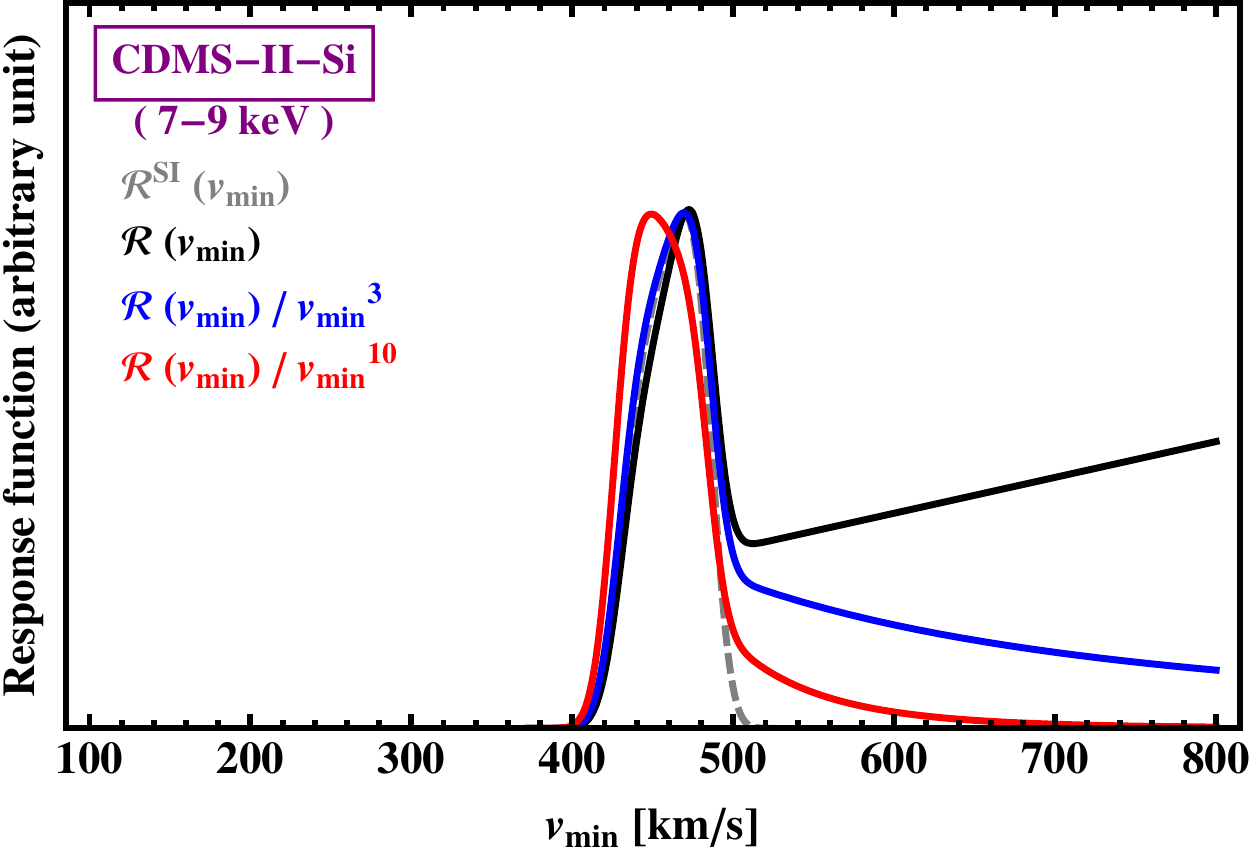}
\\
\includegraphics[width=0.49\textwidth]{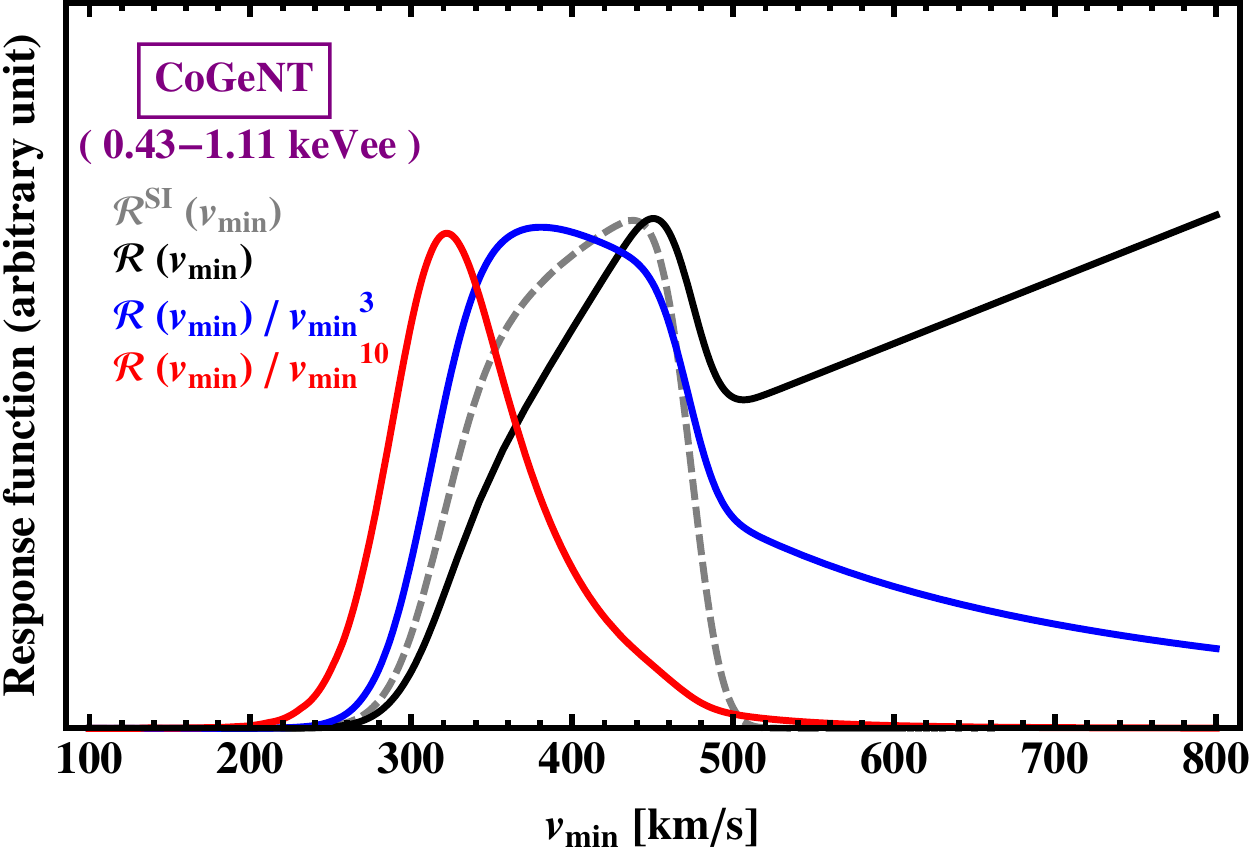}
\includegraphics[width=0.49\textwidth]{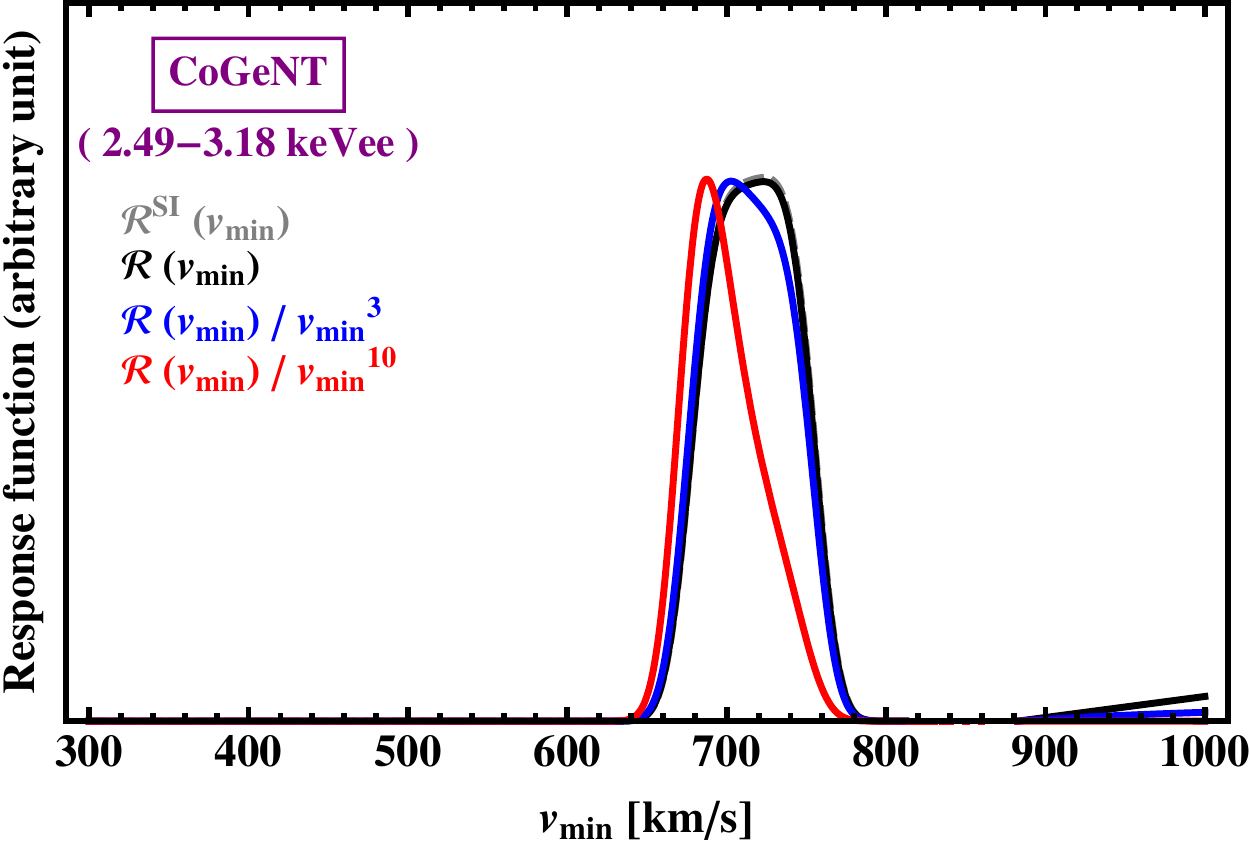}
\caption{\label{fig:responsefunctioncogent}
Response functions $\vmin^{-r} \, \eR_{[\Ed_1, \Ed_2]}(\vmin)$ with arbitrary normalization for several detected energy intervals and detectors for SI interactions (gray dashed line) and for MDM with $m = 9$ GeV/$c^2$.}\end{figure}

In fig.~\ref{fig:responsefunctioncogent} we illustrate the effect of various choices of $r$ on the response function $\vmin^{-r} \eR_{[\Ed_1, \Ed_2]}(\vmin)$ for MDM for several energy bins  and experiments: the first energy bin of DAMA/LIBRA~\cite{Bernabei:2010mq}, 2 to 2.5 keVee, the 7 to 9 keV CDMS-II used for the Si data~\cite{Agnese:2013rvf}, and the first, 0.43 to 1.11 keVee, and last, 2.49 to 3.18 keVee, of CoGeNT~\cite{Aalseth:2010vx, Aalseth:2011wp}.
We also include $\eR^{\rm SI}_{[\Ed_1, \Ed_2]}(\vmin)$ from eq.~\eqref{Resp_SI} for the standard SI interaction  (gray dashed line)  for a comparison. The normalization of each curve is arbitrary. For $r = 0$, the MDM response function is divergent and goes like $v$ at large velocities, given the $v^2$ behavior of $(v^2 \, \ud\sigma_T / \ud\ER)$ (see discussion after eq.~\eqref{eq:RT}). The divergent behavior is much more pronounced in the low-energy bins.
The choice $r = 3$ is already enough to regularize the divergent behavior, but still yields too large $\vmin$ intervals. For growing values of $r$,  the peak of the response function shifts towards low velocities (mostly in the low energy bins), due to the $\vmin^{-r}$ factor. This peak, when far from the  $\vmin$ interval where $\eR_{[\Ed_1, \Ed_2]}(\vmin)$ is non-negligible, is  unreliable as it is due to the low energy tail of the detector energy resolution function $G_T(\ER, \Ed)$, which determines the low velocity tail of $\eR_{[\Ed_1, \Ed_2]}(\vmin)$ (see eq.~\eqref{eq:HT})  and is never well known. We found the optimum $r$ value by trial an error and for MDM we find that $r = 10$  is an adequate choice (see fig.~\ref{fig:responsefunctioncogent})  to get a localized response function in $\vmin$ space without relying on how the low energy tail of the energy resolution  function is modeled. The choice of $r$ is dictated by the lowest energy bins, where the function $\vmin^{-r}$ is largest. Higher energy bins are less sensitive to the choice of $r$. 

Let us remark that  this way of comparing data is not an inherent  part of the halo-independent method but is only due to the choice of finding averages over measured energy bins to translate  putative measurements of a DM signal. So far a better way of presenting the data has not been found, and more work is necessary to make progress in this respect.

\begin{figure}[t]
\centering
\includegraphics[width=0.6\textwidth]{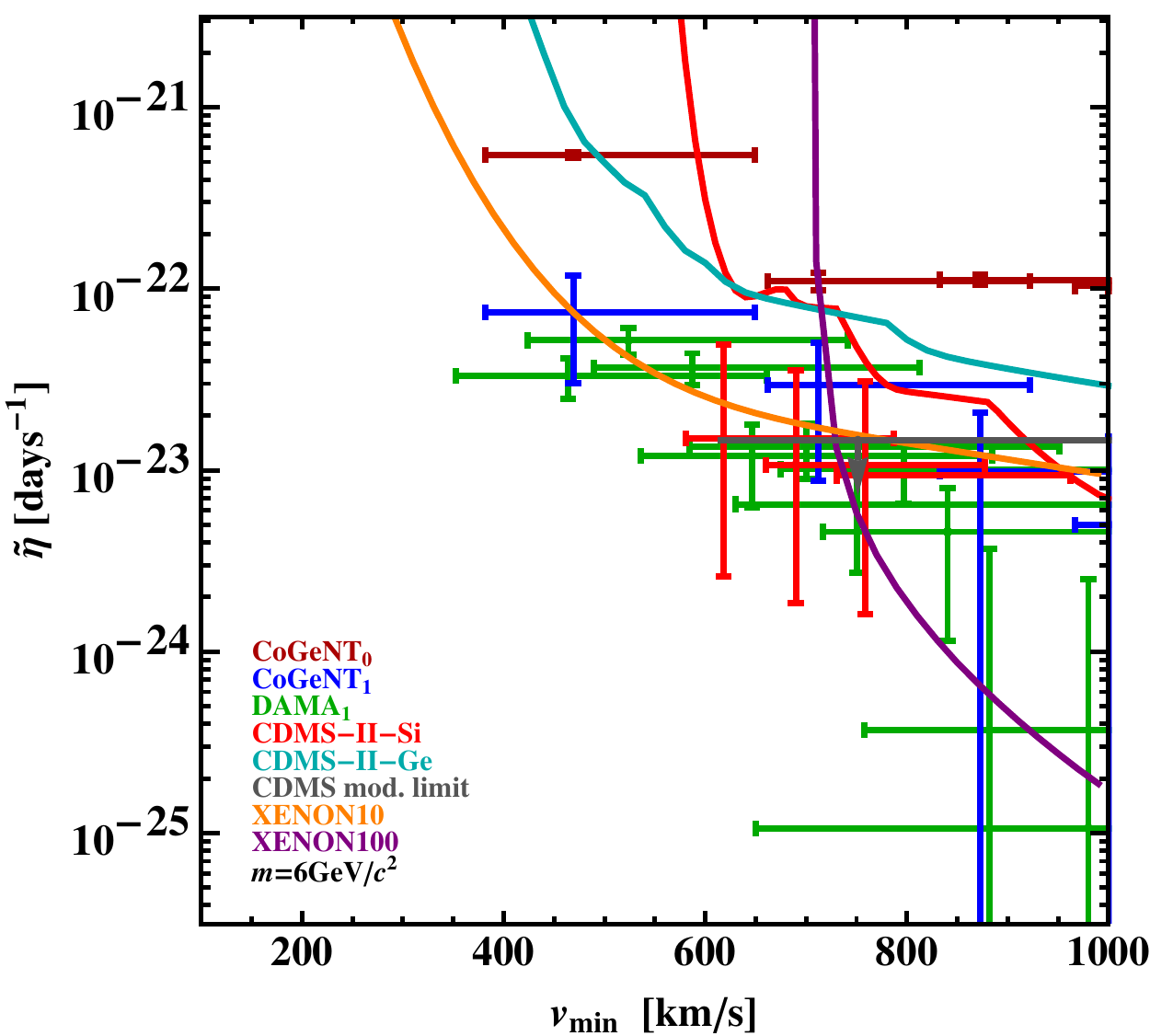}
\caption{\label{fig:2}
Measurements of and bounds on $\vmin^{-10} \overline{\vmin^{10} \tilde\eta^0(\vmin)} c^2$ and $\vmin^{-10} \overline{\vmin^{10} \tilde\eta^1(\vmin)} c^2$ for MDM of mass $m = 6$ GeV/$c^2$.}
\end{figure}

\begin{figure}[t]
\centering
\includegraphics[width=0.6\textwidth]{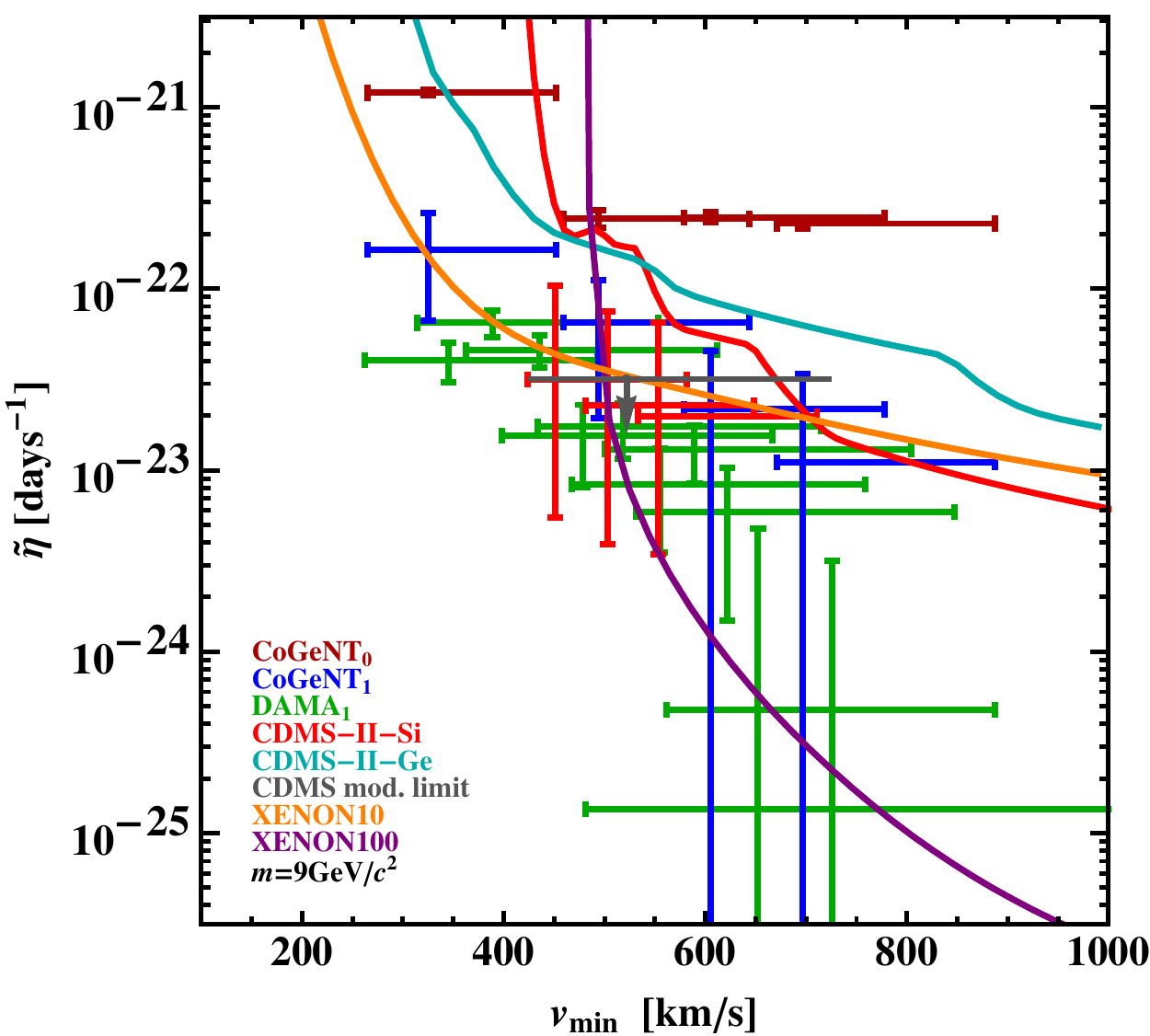}
\caption{\label{fig:3} As in fig.~\ref{fig:2} but for $m = 9$ GeV/$c^2$. All data points have moved to smaller $\vmin$ values as expected.
}
\end{figure}

\begin{figure}[t]
\centering
\includegraphics[width=0.6\textwidth]{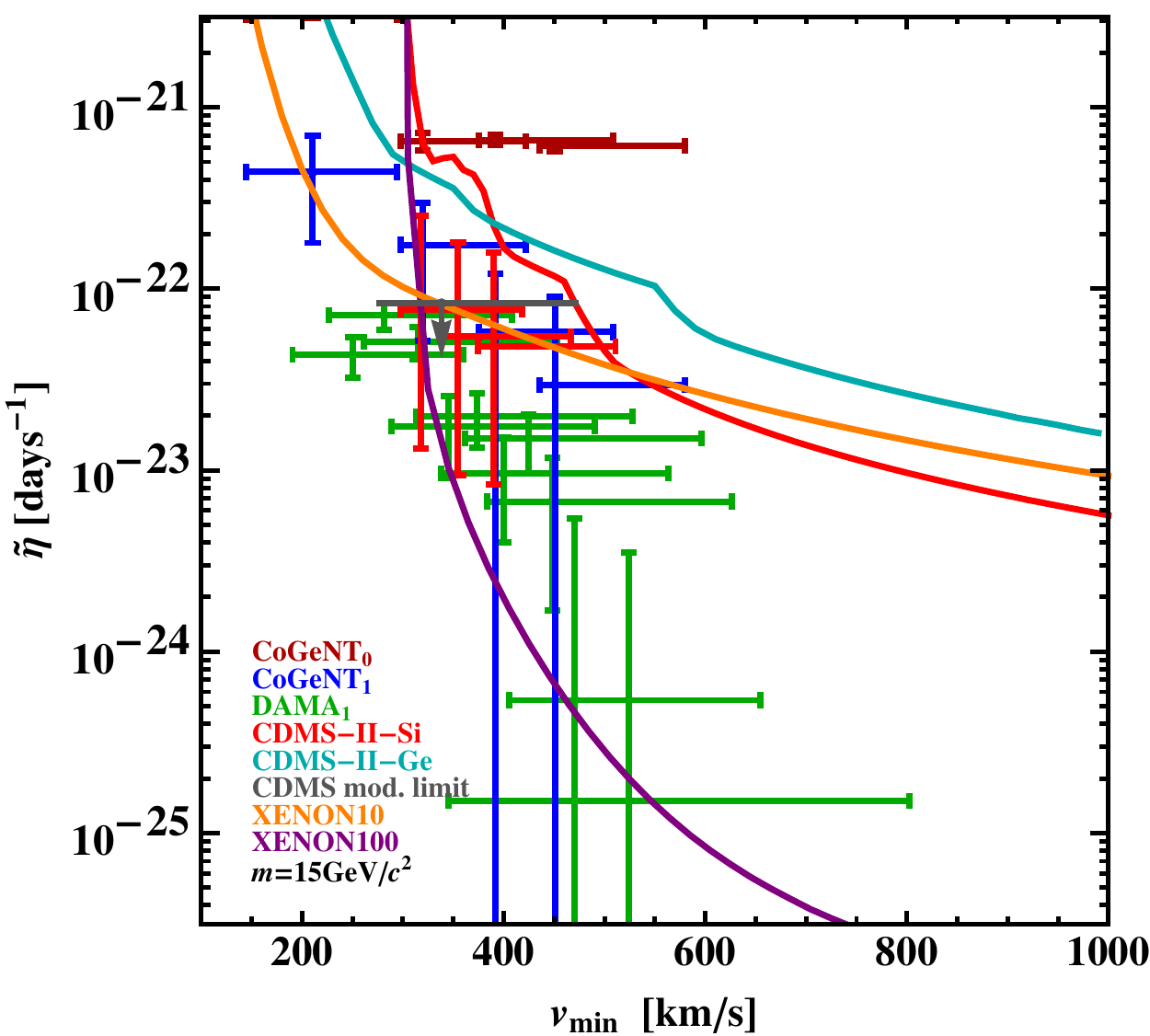}
\caption{\label{fig:4} As in fig.~\ref{fig:2} but for $m = 15$ GeV/$c^2$.
}
\end{figure}

Figs.~\ref{fig:2}, \ref{fig:3} and \ref{fig:4} show the measurements and bounds on $\vmin^{10} \tilde{\eta}^{0}(\vmin)$ and $\vmin^{10}\tilde{\eta}^{1}(\vmin)$ for a WIMP with magnetic dipole interactions and mass $m = 6$ GeV/$c^2$, $9$ GeV/$c^2$ and $15$ GeV/$c^2$ respectively. These masses are motivated by previous studies on MDM as a potential explanation for the putative DM signal found by DAMA, CoGeNT and CRESST-II (see \eg \cite{DelNobile:2012tx}). The averages (indicated by the crosses) and upper bounds are multiplied by $\vmin^{-10}$ so that the vertical axis has the usual $\tilde{\eta} c^2$ units of day$^{-1}$ and the bounds show $\tilde{\eta}^{\rm lim}(\vmin)$ (as usual for SI interactions). Figs.~\ref{fig:2}, \ref{fig:3} and \ref{fig:4} include the DAMA modulation signal (green crosses), CoGeNT modulated (blue crosses) and unmodulated signal (plus an unknown flat background, dark red horizontal lines), CDMS-II-Si unmodulated rate signal (red crosses and limit line), CDMS-II-Ge unmodulated rate limit (light blue line) and modulation bound (dark grey horizontal line with downward arrow), XENON100 limit (purple line), and XENON10 S2-only limit without $\mathcal{Q}_{\rm y}$ suppression below 1.4 keVnr (orange line). The crosses represent the averages $\overline{\vmin^{10} \tilde{\eta}^{\, i}}$  ($i=0$ for the unmodulated and $i=1$ for the modulated parts of the velocity integral) over the $\vmin$ intervals indicated by the horizontal bar of each cross, multiplied by $\vmin^{-10}$. The lines represent upper limits on $\tilde{\eta}^{0}(\vmin)$. The CDMS-II-Ge modulation limit is instead an upper limit on $\overline{\vmin^{10} \tilde{\eta}^{\, 1}}$ multiplied by $\vmin^{-10}$. 
 
The measurements and limits in figs.~\ref{fig:2}, \ref{fig:3} and \ref{fig:4} for MDM move to larger $\vmin$ values as the WIMP mass increases, as expected from the relation \eqref{vmin} between $\vmin$ and the recoil energy.  As shown in fig.~\ref{fig:2}, for a WIMP of mass $m=6$ GeV/$c^2$ the three CDMS-II-Si points are largely below the XENON10 and XENON100 upper limits, but they move progressively above them as $m$ increases to 9 GeV/$c^2$, see fig.~\ref{fig:3}, and are almost entirely excluded by them for $m=15$ GeV/$c^2$ in fig.~\ref{fig:4}. The addition of the recent LUX bound, which is not shown here, poses however a greater threat to the DM interpretation of the CDMS-II-Si excess even for very low DM masses. Moreover, the three CDMS-II-Si points overlap with or are below the CoGeNT and DAMA measurements of $\tilde{\eta}^1$, and therefore these appear incompatible with the interpretation of the CDMS-II-Si data as a measurement of $\tilde{\eta}^0$, as one usually expects that $\tilde{\eta}^0 \gg \tilde{\eta}^1$. For all three WIMP masses shown in the figures, the DAMA and CoGeNT modulation measurements seem compatible with each other. However, the upper limits on the unmodulated part of the rate imposed by XENON10 and XENON100 (plus CDMSlite and LUX, which are not shown here) reject the MDM interpretation of the DAMA and CoGeNT modulation signal, except for the lowest energy bins.

\section{Conclusions}

In order to interpret and compare data from different experiments, a model is often (if not always) needed. Concerning direct DM detection experiments, one needs to assume a model of particle interactions between DM particles and nuclei in the detectors, as well as a model for the dark halo, most notably the DM velocity distribution. For light WIMPs with $\sim 10$ GeV/$c^2$ mass, as those pointed by direct detection experiments with positive signals in the assumption of SI interactions and the SHM, the details of the high velocity region of the DM velocity distribution are crucial in comparing positive and negative results. For this reason, a framework to analyze the direct detection data independently on the properties of the DM halo is an important tool to address the compatibility of the different experiments.

In this work we have reviewed the halo-independent method to compare data from direct DM detection experiments, as introduced and developed in \cite{Fox:2010bz, Frandsen:2011gi, Gondolo:2012rs, Frandsen:2013cna, DelNobile:2013cta, DelNobile:2013cva, DelNobile:2013gba, DelNobile:2014eta}. We followed closely the treatment in \cite{DelNobile:2013cta, DelNobile:2013cva, DelNobile:2013gba}, which present the most updated analyses of DM direct detection experiment data in the context of the halo-independent method. We applied the halo-independent analysis to SI interactions with both isospin-conserving and isospin-violating couplings. In both cases the situation seems to be of disagreement between most of the experiments with positive signals (DAMA, CoGeNT, CRESST-II) and those with negative results (most notably LUX, XENON, and CDMS-II). The three CDMS-II-Si events seem however compatible with all the limits for DM with isospin-violating couplings. DAMA and CoGeNT modulation data sets seem to agree with one another, but they appear to be incompatible with the CDMS-II-Si events when these are interpreted as measurements of the unmodulated rate.

We have also shown the results of the halo-independent analysis in the assumption of WIMPs interacting with nuclei only through a magnetic dipole moment. In this case the scattering cross section has a more complicated dependence on the DM velocity, thus requiring a generalized version of the method as it was originally devised. The conclusions for this DM candidate are similar as above, with only the lowest energy data points of DAMA, CoGeNT and CDMS-II-Si lying below the exclusion bounds. The situation is somewhat better for very light WIMPs ($m \sim 6$ GeV/$c^2$), especially for the CDMS-II-Si events.

The halo-independent analysis is a promising framework to compare different direct detection experiments without making assumptions on the DM halo. This feature is highly desirable given the crucial role played by the DM velocity distribution in the galaxy in determining the total scattering rate at direct detection experiments. This analysis allows to directly compare the recoil spectra measured by different experiments in $\vmin$ space, together with bounds from null experiments. These spectra indicate the integrated DM velocity distribution $\tilde{\eta}$ favored by the experiments, as a function of $\vmin$ (see eq.~\eqref{tildeeta}).

At present this framework presents some drawbacks, which could be addressed and improved in future work. For instance, the relation between the $\tilde{\eta}$ function that one wants to fit and the observed rates is an integral equation, eq.~\eqref{R4}. So far it has been assumed that $\tilde{\eta}(\vmin, t)$ is approximately constant in any $\vmin$ interval where the response functions $\eR_{[\Ed_1, \Ed_2]}(\vmin)$ are significantly different from zero, so that it could be extracted from the integral in the form of the average in eq.~\eqref{avereta} or eqs.~\eqref{averetavr}, \eqref{averetavr2}. However, this is not necessarily a good assumption. Secondly, the degree of agreement or disagreement between two data sets can not be readily quantified (in a statistical sense) in the current halo-independent analysis. Finally, the method provides no information on the consistency of modulated and unmodulated signals, even when these are measured by the same experiment as for CoGeNT. By making some (mild) assumptions, more stringent limits on the modulated part $\tilde{\eta}^1$ can be derived from the limits on the unmodulated part of the rate \cite{Frandsen:2011gi, HerreroGarcia:2011aa, HerreroGarcia:2012fu, Bozorgnia:2013hsa}. However, with no additional assumptions the only way one can bound the modulated rate with the unmodulated rate is by imposing the most general inequality $\tilde{\eta}^0 > \tilde{\eta}^1$.

\section*{Acknowledgments}

The author thanks L.~Marcus and L.~Prosperi for useful discussions. Partial support from Department of Energy under Award Number DE-SC0009937 is also acknowledged.

\section*{Conflicts of interest}
The author declares that there is no conflict of interests regarding the publication of this article.

\end{document}